\documentclass[review]{elsarticle}

\usepackage{lineno,hyperref}
\usepackage{underscore}
\usepackage{amsfonts}
\usepackage{amsmath}
\usepackage{amssymb}
\usepackage{hyperref}
\usepackage{color}
\usepackage{graphicx}
\usepackage{amssymb}
\usepackage{amsmath}
\usepackage{graphicx}
\usepackage{dcolumn}
\usepackage{bm}
\usepackage{epsfig}
\usepackage[T1]{fontenc}
\usepackage{ae,aecompl}
\biboptions{numbers,sort&compress}
\modulolinenumbers[5]

\journal{Journal of \LaTeX\ Templates}

\begin{document}

\begin{frontmatter}

\title{Discrete Boltzmann model for implosion and explosion related compressible flow
with spherical symmetry}

\author[address1,address2]{Aiguo Xu\corref{correspondingauthor}}
\cortext[correspondingauthor]{Corresponding author}
\ead{Xu\_Aiguo@iapcm.ac.cn}
\author[address1]{Guangcai Zhang}
\author[address1,address3]{Yudong Zhang}
\author[address1]{Pei Wang}
\author[address1]{Yangjun Ying}
\address[address1]{National Key Laboratory of Computational Physics, Institute of Applied Physics and
Computational Mathematics, P. O. Box 8009-26, Beijing 100088, P.R.China}

\address[address2]{Center for Applied Physics and Technology, MOE Key Center for High Energy Density
Physics Simulations, College of Engineering, Peking University, Beijing 100871, China}

\address[address3]{Key Laboratory of Transient Physics, Nanjing University of Science and Technology, Nanjing 210094, China}

\begin{abstract}
To kinetically model implosion and explosion related phenomena, we present a theoretical framework for constructing Discrete Boltzmann Model(DBM) with spherical symmetry in spherical coordinates. To this aim, a key technique is to use \emph{local} Cartesian coordinates to describe the particle velocity in the kinetic model. Thus, the geometric effects, like the divergence and convergence, are described as a \textquotedblleft force term\textquotedblright. To better access the nonequilibrium behavior, even though the corresponding hydrodynamic model is one-dimensional, the DBM uses a Discrete Velocity Model(DVM) with 3 dimensions. A new scheme is introduced so that the DBM can use the same DVM no matter considering the extra degree of freedom or not. As an example, a DVM with 26 velocities is formulated to construct the DBM in the Navier-Stokes level. Via the DBM, one can study simultaneously the hydrodynamic and thermodynamic nonequilibrium behaviors in the implosion and explosion process which are not very close to the spherical center. The extension of current model to the multiple-relaxation-time version is straightforward.
\end{abstract}


\end{frontmatter}


\section{Introduction}

Compressible flows are ubiquitous in natural and engineering fields. The
study of compressible flow is often associated with the flight of modern
high-speed aircraft and atmospheric reentry of space-exploration vehicles;
high-Mach-number combustion system \cite{Combustion}, hydrodynamic
instabilities in inertial confinement fusion \cite{ICF}, materials under
strong shock or detonation\cite{SciChina}, etc. Common flows which are
generally in large scale and slow kinetic mode can be described by the
Navier-Stokes equations. It has been known that the Navier-Stokes model
encounters problems in describing shock wave, detonation wave, boundary
layer, micro flows, and flows with very quick kinetic modes. The physical
reasons can be understood as below. According to the Chapman-Enskog
analysis, the Navier-Stokes equations are the set of hydrodynamic equations
where only the first order Thermodynamic Non-Equilibrium (TNE) in Knudsen
number are taken into account. However, the Knudsen numbers in those small
structures and quick kinetic modes are not very small, which challenge the
validity of Navier-Stokes modeling. At the same time, in the traditional
Navier-Stokes modeling the TNE effects are coarsely described by the viscous
stress and heat flux with phenomenological constitutive relations.

In recent years, a variety of kinetic models based on Boltzmann equation are
proposed to simulate non-equilibrium flows\cite%
{2004-2009-Li-JCP,2015-Li-PAS,ZHLi2017,ZHLi2015,ZHLi2014,2014-Zhang-JFM,2015-Zhang-JCP,Zhong-PRE2017,Zhong-POF2017,TangEPL2008,TangPRE2008A,TangPRE2008B,MengJCP2011,MengPRE2011A,MengPRE2011B,MengJPCS2012,MengJFM2013,KXu2010,KXu2011,KXu2012}%
. The Discrete Boltzmann Method (DBM) \cite%
{InTech2018,2015-AG-APS,2014-AG-PRE,2015-AG-PRE,2015-AG-SM,2016-AG-PRE,2016-AG-CF,2016-AG-CF2,2017-AG-CTP,2018-AG-JMES,2018-AG-PRE,Lin2017,2018-AG-FoP,Succi-JCP}
recently developed from the Lattice Boltzmann Method (LBM) \cite%
{Succi,Succi2017,2006-Shan-JFM,2013-Zhang-JFM,Sun2009,Huang2015,Guo2016, Liu2017,Chen2015,Zhong2012,Zhang-JCP2014,Wang2010,Yeomans,XuReview2012,Watari2004}
belongs to this category. The Boltzmann equation presents values and
evolutions of all kinetic moments of the distribution function. Similar to,
but different from, the original Boltzmann equation, the DBM presents not
only values and evolutions of conserved kinetic moments (density, momentum
and energy) but also those of some nonconserved kinetic moments. The former
correspond to those described by hydrodynamic equations, the latter
complement the former in finer description of specific status and help to
understand the nonlinear constitutive relations of non-equilibrium flows\cite%
{InTech2018,2015-AG-APS}. In recent years, the DBM has brought some new
physical insights into the fundamental mechanisms of various complex flow
systems. For example, the TNE intensity has been used to discriminate the
spinodal decomposition stage and the domain growth stage in phase separation
\cite{2015-AG-SM}; the abundant TNE characteristics have been used to
distinguish and capture various interfaces \cite{2014-AG-PRE,2016-AG-PRE} in
numerical experiments, to investigate the fundamental mechanisms for entropy
increase \cite{2016-AG-CF2} in complex flows. Some of the new observations
brought by DBM, for example, the nonequilibrium fine structures of shock
waves, have been confirmed and supplemented by the results of molecular
dynamics \cite{Kang-1,Kang-2,Liu2017Recent}.

Up to now, most of DBM models for compressible fluids are in Cartesian
coordinates, except for the one in polar coordiabtes\cite{2014-AG-PRE}. In
traditional modeling the implosion and explosion processes, one-dimensional
hydrodynamic model is frequently used to describe system with spherical
symmetry and system with cylindrical symmetry with translational symmetry.
In this work we aim to construct the DBM for compressible flow systems with
spherical symmetry.

This paper is organized as below. In section II we briefly review the
kinetic and hydrodynamic models of the fluid system. In terms of their
correlations, we formulate two set of measures for the deviation of the
system from its thermodynamic equilibrium. The discrete Boltzmann models are
formulated and some numerical calculation results are shown in section III. Section IV presents the conclusion and
discussions.

\section{Fluid models}

\subsection{Kinetic model}

The Boltzmann BGK model reads%
\begin{equation}
\partial _{t}f+\mathbf{v}\cdot \nabla f=-\frac{1}{\tau }\left(
f-f^{eq}\right) \text{,}  \label{e1}
\end{equation}%
where $f=f\left( \mathbf{R}\text{, }\mathbf{v}\text{, }t\right) =f\left( x%
\text{,}y\text{,}z\text{,}v_{x}\text{,}v_{y}\text{,}v_{z}\text{,}t\right) $,
$\mathbf{R=}x\mathbf{\hat{x}}+y\mathbf{\hat{y}}+z\mathbf{\hat{z}}$ and $%
\mathbf{v=}v_{x}\mathbf{\hat{x}+}v_{y}\mathbf{\hat{y}+}v_{z}\mathbf{\hat{z}}$
in Cartesian coordinates.

\begin{figure}
  \centering
  \includegraphics[width=0.7\textwidth]{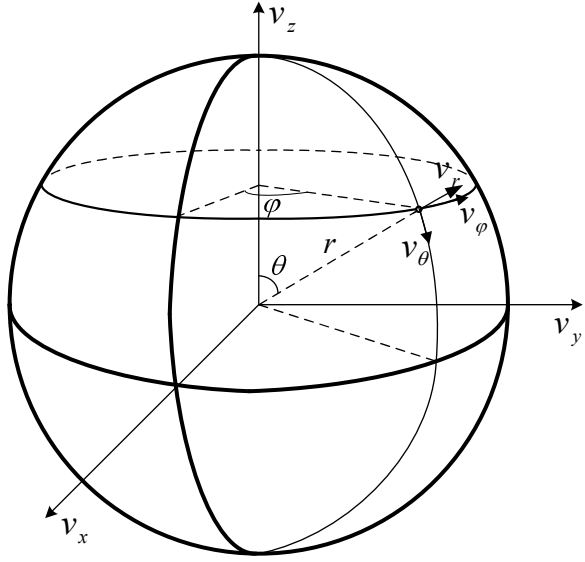}\\
  \caption{Diagrammatic drawing of spherical coordinate frame and Cartesian coordinate frame.}\label{fig0-1}
\end{figure}
In spherical coordinates, as shown in Figure \ref{fig0-1}, the position $\mathbf{R=}r\mathbf{\hat{r}}$. The three parameters $r$, $\theta $ and $\varphi $ are the radial, azimuth and zenith angle, respectively. The unit vectors, $\mathbf{\hat{r}}$, $\pmb{\hat{\theta}}$ and $\pmb{\hat{\varphi}}$, are the changing directions of
the position vector $\mathbf{R}$ along the three parameters, $r$, $\theta $
and $\varphi $, respectively, i.e.,
\begin{equation*}
d\mathbf{R}=\mathbf{\hat{r}}dr+r\pmb{\hat{\theta}}d\theta +r\sin \theta \pmb{\hat{\varphi}}d\varphi \text{.}
\end{equation*}%
Obviously, $\mathbf{\hat{r}}$, $\pmb{\hat{\theta}}$ and $\pmb{\hat{%
\varphi}}$ are orthogonal to each other and satisfy the following
relationships,
\begin{eqnarray*}
\pmb{\hat{\varphi}} &=\pmb{\hat{r}} \times \pmb{\hat{\theta}}& \mathtt{,} \\
\pmb{\hat{\theta}} &=\pmb{\hat{\varphi}} \times \pmb{\hat{r}}& \mathtt{,} \\
\mathbf{\hat{r}} &=\pmb{\hat{\theta}} \times \pmb{\hat{\varphi}}& \mathtt{.}
\end{eqnarray*}%
It is easy to find that%
\begin{eqnarray*}
d\mathbf{\hat{r}} &=&\left( \pmb{\hat{\varphi}}d\theta +\mathbf{\hat{z%
}}d\varphi \right) \times \pmb{\hat{r}}=\pmb{\hat{\theta}}d\theta +\pmb{%
\hat{\varphi}}\sin \theta d\varphi , \\
d\pmb{\hat{\theta}} &=&\left( \pmb{\hat{\varphi}}d\theta +\mathbf{%
\hat{z}}d\varphi \right) \times \pmb{\hat{\theta}}=-\pmb{\hat{r}}d\theta +%
\pmb{\hat{\varphi}}\cos \theta d\varphi , \\
d\pmb{\hat{\varphi}} &=&\left( \pmb{\hat{\varphi}}d\theta +%
\mathbf{\hat{z}}d\varphi \right) \times \boldsymbol{\hat{\varphi}}\mathbf{=-}%
\left( \mathbf{\hat{r}}\sin \theta +\pmb{\hat{\theta}}\cos \theta \right)
d\varphi .
\end{eqnarray*}

The particle velocity $\mathbf{v}$ can use the any one of the two sets of
Cartesian coordinates, $\left( \mathbf{\hat{x}}\text{, }\mathbf{\hat{y}}%
\text{, }\mathbf{\hat{z}}\right) $ and $\left( \mathbf{\hat{r}}\text{, }%
\pmb{\hat{\theta}}\text{, }\pmb{\hat{\varphi}}\right) $, as below:%
\begin{equation*}
\mathbf{v}\mathbf{=}v_{x}\mathbf{\hat{x}+}v_{y}\mathbf{\hat{y}+}v_{z}\mathbf{%
\hat{z}}
\end{equation*}%
or
\begin{equation*}
\mathbf{v}=v_{r}\mathbf{\hat{r}+}v_{\theta }\pmb{\hat{\theta}+}v_{\varphi
}\pmb{\hat{\varphi}}\text{,}
\end{equation*}%
where $v_{r}=\mathbf{v}\cdot \mathbf{\hat{r}}$, $v_{\theta }=\mathbf{v%
}\cdot \pmb{\hat{\theta}}$ and $v_{\varphi }=\mathbf{v}\cdot \pmb{%
\hat{\varphi}}$ are the projections of $\mathbf{v}$ in the $\mathbf{\hat{r}}$%
, $\pmb{\hat{\theta}}$, and $\pmb{\hat{\varphi}}$ directions,
respectively. Correspondingly, the distribution function $f$ can also be
described in two different forms,
\begin{equation}
f=f\left( r\text{, }\theta \text{, }\varphi \text{, }v_{x}\text{, }v_{y}%
\text{, }v_{z}\text{, }t\right)  \label{eq6}
\end{equation}%
or
\begin{equation}
f=f\left( r\text{, }\theta \text{, }\varphi \text{, }v_{r}\text{, }v_{\theta
}\text{, }v_{\varphi }\text{, }t\right) \text{.}  \label{eq7}
\end{equation}%
In this work we will use the latter form, Eq. (\ref{eq7}). Under the
definition (\ref{eq7}), it should be stressed that, when calculate the
spatial derivative, $\nabla f$, even though the particle velocity $\mathbf{v}
$ is fixed, its three components vary with the position $\mathbf{R}$, i.e., $%
v_{r}=v_{r}\left( \mathbf{R}\right) $, $v_{\theta }=v_{\theta }\left(
\mathbf{R}\right) $, and $v_{\varphi }=v_{\varphi }\left( \mathbf{R}\right) $%
. We use the symbol \textquotedblleft $\nabla |_{\mathbf{v}}$%
\textquotedblright\ to replace \textquotedblleft $\nabla $%
\textquotedblright\ in Eq. (\ref{e1}) to stress that $\mathbf{v}$ is fixed
when calculating the spatial derivatves. Thus, Eq. (\ref{e1}) is rewritten
as
\begin{equation}
\partial _{t}f+\mathbf{v}\cdot \nabla |_{\mathbf{v}}f=-\frac{1}{\tau }\left(
f-f^{eq}\right) .  \label{eq8}
\end{equation}%
According to the definition (\ref{eq7}),
\begin{equation}
\nabla |_{\mathbf{v}}f=\nabla |_{v_{r},v_{\theta },v_{\varphi }}\text{ }%
f+\nabla v_{r}\frac{\partial f}{\partial v_{r}}+\nabla v_{\theta }\frac{%
\partial f}{\partial v_{\theta }}+\nabla v_{\varphi }\frac{\partial f}{%
\partial v_{\varphi }}\text{,}  \label{eq9}
\end{equation}%
where
\begin{equation}
\nabla v_{r}=\frac{v_{\theta }}{r}\mathbf{\hat{\theta}}+\frac{v_{\varphi }}{r%
}\mathbf{\hat{\varphi}}\mathtt{,}  \label{eq10}
\end{equation}
\begin{equation}
\nabla v_{\theta }=-\frac{v_{r}}{r}\mathbf{\hat{\theta}+}\frac{v_{\varphi
}\cos \theta }{r\sin \theta }\mathbf{\hat{\varphi}}\text{,}  \label{eq11}
\end{equation}%
\begin{equation}
\nabla v_{\varphi }=-\left( \frac{v_{r}}{r}\mathbf{+}\frac{v_{\theta }\cos
\theta }{r\sin \theta }\right) \mathbf{\hat{\varphi}}\text{.}  \label{eq12}
\end{equation}%
Substituting Eqs. (\ref{eq9})-(\ref{eq12}) into Eq. (\ref{eq8}) gives the
Boltzmann equation in spherical coordinates,
\begin{eqnarray}
&&\partial _{t}f+\mathbf{v}\cdot \nabla |_{v_{r},v_{\theta },v_{\varphi }}%
\text{ }f+\frac{v_{\theta }^{2}+v_{\varphi }^{2}}{r}\frac{\partial f}{%
\partial v_{r}}+\left( \frac{v_{\varphi }^{2}\cos \theta }{r\sin \theta }-%
\frac{v_{r}v_{\theta }}{r}\right) \frac{\partial f}{\partial v_{\theta }}%
-\left( \frac{v_{r}v_{\varphi }}{r}+\frac{v_{\varphi }v_{\theta }\cos \theta
}{r\sin \theta }\right) \frac{\partial f}{\partial v_{\varphi }}  \notag \\
&=&-\frac{1}{\tau }\left( f-f^{eq}\right) .  \label{eq14}
\end{eqnarray}%
For macroscopic system with spherical symmetry, the distribution function $f$
does not depend on the angles $\theta $ and $\varphi $, i.e.,
\begin{equation}
f=f\left( r\text{, }v_{r}\text{, }v_{\theta }\text{, }v_{\varphi }\text{, }%
t\right) ;  \label{eq15}
\end{equation}%
and $f$ is invariant under the rotation in the subspace of $\left( v_{\theta
}\text{, }v_{\varphi }\right) $, i.e.,
\begin{equation}
f\left( r\text{, }v_{r}\text{, }v_{\theta }\text{, }v_{\varphi }\text{, }%
t\right) =f\left( r\text{, }v_{r}\text{, }v_{\theta }^{\prime }\text{, }%
v_{\varphi }^{\prime }\text{, }t\right)  \label{eq16}
\end{equation}%
when
\begin{equation}
v_{\theta }^{2}+v_{\varphi }^{2}=v_{\theta }^{\prime 2}+v_{\varphi }^{\prime
2}\text{.}  \label{eq17}
\end{equation}%
Now, we take $v_{\theta}^{\prime}= v_{\theta} + dv_{\theta}$ and $v_{\phi}^{\prime}= v_{\phi} + dv_{\phi}$. From Eq.(\ref{eq16}) we can get
\begin{equation}
f\left( r\text{, }v_{r}\text{, }v_{\theta }^{\prime }\text{, }%
v_{\varphi }^{\prime }\text{, }t\right) = f\left( r\text{, }v_{r}\text{, }v_{\theta }\text{, }v_{\varphi }\text{, }%
t\right) + \frac{\partial f}{\partial v_{\theta }}dv_{\theta }+ \frac{\partial f}{\partial v_{\phi }}dv_{\phi } \text{,}\label{eq17-2}
\end{equation}%
which gives
\begin{equation}
\frac{\partial f}{\partial v_{\theta }}dv_{\theta }+ \frac{\partial f}{\partial v_{\phi }}dv_{\phi }=0\text{.}  \label{eq18}
\end{equation}%
From Eq.(\ref{eq17}) we can get
\begin{equation}
\frac{dv_{\theta}}{v_{\phi}}= -\frac{dv_{\phi}}{v_{\theta}} \text{.}  \label{eq17-1}
\end{equation}%
So,
\begin{equation}
\left( v_{\theta }\frac{\partial f}{\partial v_{\varphi }}-v_{\varphi }\frac{%
\partial f}{\partial v_{\theta }}\right) =0\text{.}  \label{eq18}
\end{equation}%
Using the conditions, (\ref{eq15}) and (\ref{eq18}), in Eq. (\ref{eq14})
gives
\begin{equation}
\partial _{t}f+\mathbf{v}\cdot \nabla |_{v_{r},v_{\theta },v_{\varphi }}%
\text{ }f+\left( \frac{v_{\theta }^{2}+v_{\varphi }^{2}}{r}\frac{\partial f}{%
\partial v_{r}}-\frac{v_{r}v_{\theta }}{r}\frac{\partial f}{\partial
v_{\theta }}-\frac{v_{r}v_{\varphi }}{r}\frac{\partial f}{\partial
v_{\varphi }}\right) =-\frac{1}{\tau }\left( f-f^{eq}\right) .  \label{eq19}
\end{equation}%
It is clear that the term%
\begin{equation*}
\left( \frac{v_{\theta }^{2}+v_{\varphi }^{2}}{r}\frac{\partial f}{\partial
v_{r}}-\frac{v_{r}v_{\theta }}{r}\frac{\partial f}{\partial v_{\theta }}-%
\frac{v_{r}v_{\varphi }}{r}\frac{\partial f}{\partial v_{\varphi }}\right)
\end{equation*}%
plays a similar role of the \textquotedblleft force term\textquotedblright\
in the Boltzmann equation in Cartesian coordinates. It creates the
divergence or convergence effects in the flow system.
As the first step, here we consider the simplest case where the
thermodynamic nonequilibrium effects resulted from the pure geometric effects are much weaker
 than those resulted from other kinds of contributions.
In such a case, we can use the approximation, $f=f^{eq}$, when calculating the ``force term".
Such a treatment is reasonable when the flow behaviors under consideration are not close to the spherical center.
If further use the macroscopic condition, $%
u_{\theta }=u_{\varphi }=0$, for spherical symmetry, the final Boltzmann
equation for the flow system with spherical symmetry becomes,%
\begin{equation}
\partial _{t}f+v_{r}\frac{\partial f}{\partial r}+\left[ \frac{%
v_{r}v_{\theta }^{2}}{rT}+\frac{v_{r}v_{\varphi }^{2}}{rT}-\frac{\left(
v_{\theta }^{2}+v_{\varphi }^{2}\right) \left( v_{r}-u_{r}\right) }{rT}%
\right] f^{eq}=-\frac{1}{\tau }\left( f-f^{eq}\right) .  \label{eq22}
\end{equation}

\subsection{Hydrodynamic model}

The Navier-Stokes equations in Cartesian coordinates read
\begin{subequations}
\begin{eqnarray}
\frac{\partial \rho }{\partial t}+\frac{\partial (\rho u_{\alpha })}{%
\partial x_{\alpha }} &=&0\text{,}  \label{HydroCart1} \\
\frac{\partial (\rho u_{\alpha })}{\partial t}+\frac{\partial (\rho
u_{\alpha }u_{\beta })}{\partial x_{\beta }}+\frac{\partial P\delta _{\alpha
\beta }}{\partial x_{\beta }} &=&\frac{\partial \sigma _{\alpha \beta }}{%
\partial x_{\beta }}\text{,}  \label{HydroCart2} \\
\frac{\partial }{\partial t}(\rho e+\frac{1}{2}\rho u^{2})+\frac{\partial }{%
\partial x_{\alpha }}[u_{\alpha }(\rho e+\frac{1}{2}\rho u^{2}+P)] &=&\frac{%
\partial }{\partial x_{\alpha }}\left[ Q_{\alpha }+u_{\beta }\sigma
_{\alpha \beta }\right] \text{.}  \label{HydroCart3}
\end{eqnarray}%
In the left-hand side of Eqs. (\ref{HydroCart1})- (\ref{HydroCart3}), $\rho $%
, $\mathbf{u}$, $P$ $=\rho RT$, $\rho e=\left( D+n\right) RT/2$, $T$, $%
\gamma =\left( D+n+2\right) /\left( D+n\right) $ are the hydrodynamic
density, flow velocity, pressure, internal energy, temperature and
specific-heat ratio, respectively. $\alpha =x$,$y$,or $z$. $u^{2}=\mathbf{u}%
\cdot \mathbf{u}$. $D$ is the space dimension and $n$ is the number of extra
degrees of freedom. In the right-hand side of Eqs. (\ref{HydroCart2})- (\ref%
{HydroCart3}),
\end{subequations}
\begin{equation}
\sigma _{\alpha \beta }=\mu \left[ \frac{\partial u_{\beta }}{\partial
r_{\alpha }}+\frac{\partial u_{\alpha }}{\partial r_{\beta }}-\left( \frac{2%
}{D}-\lambda \right) \frac{\partial u_{\gamma }}{\partial r_{\gamma }}\delta
_{\alpha \beta }\right]
\end{equation}%
is the viscous stress and
\begin{equation}
Q_{\alpha }=\kappa\frac{\partial e}{\partial r_{\alpha }}
\end{equation}%
is the heat flux. The two parameters, $\mu $ and $\lambda $ are coefficient of viscosity
and the parameter, $\kappa$ is the coefficient of heat conductivity. $e=DRT/2$ is the
translational internal energy. It is clear that $P=2\rho e/D$. When the
viscosities and heat conductivity vanish, the hydrodynamic equations, (\ref%
{HydroCart1}) - (\ref{HydroCart3}), become the Euler equations.

\subsection{From kinetic to hydrodynamic model}

The Chapman-Enskog multiscale analysis shows that, \bigskip the
Navier-Stokes equations, (\ref{HydroCart1}) - (\ref{HydroCart3}), are the
set of hydrodynamic equations from the Boltzmann BGK equation, (\ref{e1}),
when only the first order terms in Knudsen number are taken into account.
Among the infinite number of velocity kinetic moment relations of $f^{eq}$,
only the following seven,
\begin{subequations}
\begin{equation}
\int \int f^{eq}d\mathbf{v}d\boldsymbol{\eta}=\rho \text{,}  \label{Mr1}
\end{equation}%
\begin{equation}
\int \int f^{eq}v_{\alpha }d\mathbf{v}d\boldsymbol{\eta }=\rho u_{\alpha }\text{,%
}  \label{Mr2}
\end{equation}%
\begin{equation}
\int \int f^{eq}\left( v^{2}+\eta ^{2}\right) d\mathbf{v}d\boldsymbol{\eta }%
=2\rho \left( \frac{D+n}{D}e+\frac{u^{2}}{2}\right) \text{,}  \label{Mr3}
\end{equation}%
\begin{equation}
\int \int f^{eq}v_{\alpha }v_{\beta }d\mathbf{v}d\boldsymbol{\eta }=P\delta
_{\alpha \beta }+\rho u_{\alpha }u_{\beta }\text{,}  \label{Mr4}
\end{equation}%
\begin{equation}
\int \int f^{eq}\left( v^{2}+\eta ^{2}\right) v_{\alpha }d\mathbf{v}d\boldsymbol{%
\eta }=2\rho \left( \frac{D+n+2}{D}e+\frac{u^{2}}{2}\right) u_{\alpha }\text{%
,}  \label{Mr5}
\end{equation}%
\begin{equation}
\int \int f^{eq}v_{\alpha }v_{\beta }v_{\chi }d\mathbf{v}d\boldsymbol{\eta }%
=\rho RT\left( u_{\alpha }\delta _{\beta \chi }+u_{\beta }\delta _{\alpha
\chi }+u_{\chi }\delta _{\alpha \beta }\right) +\rho u_{\alpha }u_{\beta
}u_{\chi }\text{,}  \label{Mr6}
\end{equation}%
\begin{equation}
\int \int f^{eq}\left( v^{2}+\eta ^{2}\right) v_{\alpha }v_{\beta }d\mathbf{v%
}d\boldsymbol{\eta }=\frac{4}{D}\rho e\left( \frac{D+n+2}{D}e+\frac{u^{2}}{2}%
\right) \delta _{\alpha \beta }+2\rho u_{\alpha }u_{\beta }\left( \frac{D+n+4%
}{D}e+\frac{u^{2}}{2}\right) \text{,}  \label{Mr7}
\end{equation}%
\end{subequations}
are needed in the Chapman-Enskog analysis, where $\boldsymbol{\eta }$ is a
parameter describing the fluctuating velocity in the $n$ extra degrees of
freedom, $\eta ^{2}=\boldsymbol{\eta }\cdot \boldsymbol{\eta }$, and
\begin{equation}\label{feq}
f^{eq}\left( \mathbf{v}\right) =\rho \left( \frac{1}{2\pi RT}\right)
^{(D+n)/2}\exp \left[ -\frac{\left( \mathbf{v}-\mathbf{u}\right) ^{2}+\eta^2}{2RT}%
\right] \text{.}
\end{equation}
 The Chapman-Enskog
analysis gives the following constitutive relations for the Navier-Stokes
model, (\ref{HydroCart1}) - (\ref{HydroCart3}).

\begin{equation*}
\mu =\frac{2}{D}\rho e\tau \text{,}\lambda =\frac{D+n-2}{D+n}\text{,}k=\frac{%
2\left( D+n+2\right) }{D\left( D+n\right) }\rho e\tau .
\end{equation*}

Under the spherical symmetry, the Navier-Stokes equations read
\begin{equation*}
{{\partial }_{t}}\rho +({{\partial }_{r}}+\frac{2}{r})\left( \rho u\right) =0%
\text{,}
\end{equation*}%
\begin{equation*}
{{\partial }_{t}}u+u{{\partial }_{r}}u+\frac{1}{\rho }{{\partial }_{r}}p=%
\frac{4}{\rho r}\mu ({{\partial }_{r}}u-\frac{u}{r})-\frac{1}{\rho }{{%
\partial }_{r}}\left[ \mu (1-\lambda )({{\partial }_{r}}+\frac{2}{r})u-2\mu {%
{\partial }_{r}}u\right] \text{,}
\end{equation*}%
\begin{eqnarray}
&&{{\partial }_{t}}e+u{{\partial }_{r}}e+\frac{p}{\rho }\left[ ({{\partial }%
_{r}}+\frac{2}{r})u\right]   \notag \\
&=&\frac{1}{\rho }({{\partial }_{r}}+\frac{2}{r})(k{{\partial }_{r}}T)+\frac{%
2\mu }{\rho }\left[ {{({{\partial }_{r}}u)}^{2}}+2{{(\frac{u}{r})}^{2}}%
\right] -\frac{1}{\rho }\mu (1-\lambda )\left[ ({{\partial }_{r}}+\frac{2}{r}%
)u\right] ^{2}\text{,}
\end{eqnarray}%
which can be recovered from the kinetic model (\ref{eq22}). It should be
stressed that the set of kinetic moment relations,(\ref{Mr1})-(\ref{Mr7}),
keep exactly the same form when being transfered from the $\left( v_{x}\text{%
,}v_{y}\text{,}v_{z}\right) $ coordiantes to the $\left( v_{r}\text{,}%
v_{\theta }\text{,}v_{\varphi }\right) $ coordiantes, in this work, when
using velocity kinetic moment relations, $v_{\alpha }=v_{r}$,$v_{\theta }$,$%
v_{\varphi }$.

\subsection{Measurements of nonequilibrium effects}

The Chapman-Enskog multiscale analysis tells that, as the simplest
hydrodynamic model of fluid system, the Euler equations ignore completely
the Thermodynamic Non-Equilibrium (TNE) behavior. The Navier-Stokes equations
describe the TNE behavior via the terms in viscosity and heat conductivity.
The Euler model works successfully when we consider the fluid system in a
time scale $t_{0}$ which is large enough compared with thermodynamic
relaxation time $\tau $. Besides the normal high speed compressible flows,
the Euler model works also for solid materials under strong shock. From the
mechanical side, compared with the shocking strength, the material strength,
for example, the yield, and viscous stress are negligible. Consequently, the
Euler model works better with increasing the shock strength. From the side
of time scales, when we study the shocking procedure, the time scale $t_{0} $
used is generally small enough compared with the time scale $t_{h}$ for heat
conduction and large enough compared with the thermodynamic relaxation time $%
\tau $. In other words, during the time interval under investigation, the
heat conduction does not have time to occur significantly and consequently
its effects are negligible. For the objective system where the thermodynamic
relaxation time $\tau $ is fixed, if we decreases the observing time scale $%
t_{0}$, we find more TNE effects. The Boltzmann kinetic model can be used to
investigate both the hydrodynamic and thermodynamic behaviors.

Following the seven moment relations, (\ref{Mr1})-(\ref{Mr7}), used in
recovering the Navier-Stokes equations, we define the following moments,
\begin{subequations}
\begin{equation}
\mathbf{M}_{0}^{\ast }(f\text{,}\mathbf{v})=\int \int f\text{ }d\mathbf{v}d\boldsymbol{\eta}\text{,}  \label{Meq1}
\end{equation}%
\begin{equation}
\mathbf{M}_{1}^{\ast }(f\text{,}\mathbf{v})=\int \int f\text{ }\mathbf{v}d%
\mathbf{v}d\boldsymbol{\eta}\text{,}  \label{Meq2}
\end{equation}%
\begin{equation}
\mathbf{M}_{2,0}^{\ast }(f\text{,}\mathbf{v})=\int \int f\text{ }\left(
\mathbf{v}\cdot \mathbf{v}+\eta ^{2}\right) d\mathbf{v}d\boldsymbol{\eta}\text{,%
}  \label{Meq3}
\end{equation}%
\begin{equation}
\mathbf{M}_{2}^{\ast }(f\text{,}\mathbf{v})=\int \int f\text{ }\mathbf{vv}d%
\mathbf{v}d\boldsymbol{\eta}\text{,}  \label{Meq4}
\end{equation}%
\begin{equation}
\mathbf{M}_{3,1}^{\ast }(f\text{,}\mathbf{v})=\int \int f\text{ }\left(
\mathbf{v}\cdot \mathbf{v}+\eta ^{2}\right) \mathbf{v}d\mathbf{v}d\boldsymbol{\eta}\text{,}  \label{Meq5}
\end{equation}%
\begin{equation}
\mathbf{M}_{3}^{\ast }(f\text{,}\mathbf{v})=\int \int f\text{ }\mathbf{vvv}d%
\mathbf{v}d\boldsymbol{\eta}\text{,}  \label{Meq6}
\end{equation}%
\begin{equation}
\mathbf{M}_{4,2}^{\ast }(f\text{,}\mathbf{v})=\int \int f\text{ }\left(
\mathbf{v}\cdot \mathbf{v}+\eta ^{2}\right) \mathbf{vv}d\mathbf{v}d\boldsymbol{\eta}\text{,}  \label{Meq7}
\end{equation}
where $\mathbf{M}_{n}^{\ast }$ means a $n$-th order tensor and $\mathbf{M}%
_{m,n}^{\ast }$ means a $n$-th-order tensor contracted from a $m$-th order
tensor. For the case of central moments, the variable $\mathbf{v}$ is
replaced with $\mathbf{v}^{\ast }=\left( \mathbf{v}-\mathbf{u}\right) $. It
is clear $\mathbf{M}_{0}^{\ast }$ and $\mathbf{M}_{2,0}^{\ast }$ are
scalars. Each of them has only $1$ component. $\mathbf{M}_{1}^{\ast }$ and $%
\mathbf{M}_{3,1}^{\ast }$ are vectors. Each of them has $2$ independent
components in $2$-dimensional case or $3$ independent components in $3$%
-dimensional case. $\mathbf{M}_{2}^{\ast }$ and $\mathbf{M}_{4,2}^{\ast }$
are 2nd order tensors. Each of them has $3$ independent components in $2$%
-dimensional case or $6$ independent components in $3$-dimensional case. $%
\mathbf{M}_{3}^{\ast }$ is 3rd tensor and has $4$ independent components in $%
2$-dimensional case or $10$ independent components in $3$-dimensional case.
Therefore, the constraints, (\ref{Mr1}) - (\ref{Mr7}), are in fact $16$
linear equations in $f^{eq}$ in $2$-dimensional case and $30$ linear
equations in $f^{eq}$ in $3$-dimensional case. We further define
\end{subequations}
\begin{equation}
\boldsymbol{\Delta }_{m,n}^{\ast }\left( \mathbf{v}\right) =\mathbf{M}%
_{m,n}^{\ast }(f\text{,}\mathbf{v})-\mathbf{M}_{m,n}^{\ast }(f^{eq}\text{,}%
\mathbf{v})\text{.}  \label{Deq1}
\end{equation}%
It is clear that $\boldsymbol{\Delta }_{0}^{\ast }\left( \mathbf{v}\right) =%
\mathbf{0}$, $\boldsymbol{\Delta }_{1}^{\ast }\left( \mathbf{v}\right) =%
\mathbf{0}$ and $\boldsymbol{\Delta }_{2,0}^{\ast }\left( \mathbf{v}\right) =%
\mathbf{0}$, which is due to the mass, momentum and energy conservations.
Except for the three, the quantity $\boldsymbol{\Delta }_{m,n}^{\ast }\left(
\mathbf{v}\right) $ works as a measure for the deviation of the system from
its thermodynamic equilibrium. The information of flow velocity $\mathbf{u}$
is taken into account in the definition (\ref{Deq1}). Similarly,
\begin{equation}
\boldsymbol{\Delta }_{m,n}^{\ast }\left( \mathbf{v}^{\ast }\right) =\mathbf{M%
}_{m,n}^{\ast }(f\text{,}\mathbf{v}^{\ast })-\mathbf{M}_{m,n}^{\ast }(f^{eq}%
\text{,}\mathbf{v}^{\ast })\text{.}  \label{Deq2}
\end{equation}%
Except for $\boldsymbol{\Delta }_{0}^{\ast }\left( \mathbf{v}^{\ast }\right)
$, $\boldsymbol{\Delta }_{1}^{\ast }\left( \mathbf{v}^{\ast }\right) $ and $%
\boldsymbol{\Delta }_{2,0}^{\ast }\left( \mathbf{v}^{\ast }\right) $, the
quantity $\boldsymbol{\Delta }_{m,n}^{\ast }\left( \mathbf{v}^{\ast }\right)
$ works as a measure for the deviation of the system from its thermodynamic
equilibrium, where only the thermal fluctuations of the molecules are
considered.

\section{Discrete Boltzmann models for systems with spherical symmetry}

There are two key techniques in constructing DBM with force terms. The first
is to calculate the velocity derivative of $f$, $\partial f/\partial \mathbf{%
v}$, before discretzing the particle velocity space. As the first step, one
can consider the case where $f$ can be approximated by $f^{eq}$ in the force
term. The second is to write the discrete equilibrium distribution function,$%
f^{eq}_{i}$, as a function of the discrete velocities, where $i$ is the
index of the discrete velocity.

For constructing the DBM for systems with spherical symmetry, we use Eq. (%
\ref{eq22}). We have
\begin{equation}
\partial _{t}f_{i}+v_{ir}\frac{\partial f_{i}}{\partial r}+\left[ \frac{%
v_{ir}v_{i\theta }^{2}}{rT}+\frac{v_{ir}v_{i\varphi }^{2}}{rT}-\frac{\left(
v_{i\theta }^{2}+v_{i\varphi }^{2}\right) \left( v_{ir}-u\right) }{rT}\right]
f_{i}^{eq}=-\frac{1}{\tau }\left( f_{i}-f_{i}^{eq}\right) .  \label{DBS1}
\end{equation}%
where $f_{i}$ ($f_{i}^{eq}$) is the discrete (equilibrium) distribution
function; $v_{i}$ is the $i$-th discrete velocity, $i=1$, $...$, $N$; $N$ is
the total number of the discrete velocity.

The fundamental requirement for a DBM is that it should recover the same set
of hydrodynamic equations as those given by the original continuous
Boltzmann equation. The Chapman-Enskog multiscale analysis shows that, only
if the seven moment relations, (\ref{Mr1})-(\ref{Mr7}), can be calcualted
equally in the summation form as below,

\begin{subequations}
\begin{equation}
\rho =\sum_{i=1}^{N}f_{i}^{eq}=\sum_{i=1}^{N}f_{i}\text{,}  \label{m1}
\end{equation}%
\begin{equation}
\rho u_{\alpha }=\sum_{i=1}^{N}f_{i}^{eq}v_{i\alpha
}=\sum_{i=1}^{N}f_{i}v_{i\alpha }\text{,}  \label{m2}
\end{equation}%
\begin{equation}
2\rho \left( \frac{D+n}{D}e+\frac{u^{2}}{2}\right)
=\sum_{i=1}^{N}f_{i}^{eq}(v_{i}^{2}+\eta
_{i}^{2})=\sum_{i=1}^{N}f_{i}(v_{i}^{2}+\eta _{i}^{2})\text{,}  \label{m3}
\end{equation}%
\begin{equation}
P\delta _{\alpha \beta }+\rho u_{\alpha }u_{\beta
}=\sum_{i=1}^{N}f_{i}^{eq}v_{i\alpha }v_{i\beta }\text{,}  \label{m4}
\end{equation}%
\begin{equation}
2\rho \left( \frac{D+n+2}{D}e+\frac{u^{2}}{2}\right) u_{\alpha
}=\sum_{i=1}^{N}f_{i}^{eq}(v_{i}^{2}+\eta _{i}^{2})v_{i\alpha }\text{,}
\label{m5}
\end{equation}
\begin{equation}
\rho RT\left( u_{\alpha }\delta _{\beta \chi }+u_{\beta }\delta _{\alpha
\chi }+u_{\chi }\delta _{\alpha \beta }\right) +\rho u_{\alpha }u_{\beta
}u_{\chi }=\sum f_{i}^{eq}v_{i\alpha }v_{i\beta }v_{i\chi },  \label{m6}
\end{equation}%
\begin{equation}
\frac{4}{D}\rho e\left( \frac{D+n+2}{D}e+\frac{u^{2}}{2}\right) \delta
_{\alpha \beta }+2\rho u_{\alpha }u_{\beta }\left( \frac{D+n+4}{D}e+\frac{%
u^{2}}{2}\right) =\sum f_{i}^{eq}\left( v_{i}^{2}+\eta _{i}^{2}\right)
v_{i\alpha }v_{i\beta }\text{,}  \label{meq7}
\end{equation}%
\end{subequations}
Navier-Stokes model, (\ref{HydroCart1}) - (\ref{HydroCart3}), can be
recovered from the discrete Boltzmann model,(\ref{DBS1}). Following the same
idea as in the definitions, (\ref{Meq1}) - (\ref{Meq7}), we define the
following moments of the discrete distribution function $f_{i}$,

\begin{subequations}
\begin{equation}
\mathbf{M}_{0}(f_{i}\text{,}\mathbf{v}_{i})=\sum_{i=1}^{N}f_{i}\text{,}
\label{Me1}
\end{equation}%
\begin{equation}
\mathbf{M}_{1}(f_{i}\text{,}\mathbf{v}_{i})=\sum_{i=1}^{N}f_{i}\mathbf{v}_{i}%
\text{,}  \label{Me2}
\end{equation}%
\begin{equation}
\mathbf{M}_{2,0}(f_{i}\text{,}\mathbf{v}_{i})=\sum_{i=1}^{N}f_{i}(\mathbf{v}%
_{i}\cdot \mathbf{v}_{i}+\eta _{i}^{2})\text{,}  \label{Me3}
\end{equation}%
\begin{equation}
\mathbf{M}_{2}(f_{i}\text{,}\mathbf{v}_{i})=\sum_{i=1}^{N}f_{i}\mathbf{v}_{i}%
\mathbf{v}_{i}\text{,}  \label{Me4}
\end{equation}%
\begin{equation}
\mathbf{M}_{3,1}(f_{i}\text{,}\mathbf{v}_{i})=\sum_{i=1}^{N}f_{i}(\mathbf{v}%
_{i}\cdot \mathbf{v}_{i}+\eta _{i}^{2})\mathbf{v}_{i}\text{,}  \label{Me5}
\end{equation}%
\begin{equation}
\mathbf{M}_{3}(f_{i}\text{,}\mathbf{v}_{i})=\sum f_{i}\mathbf{v}_{i}\mathbf{v%
}_{i}\mathbf{v}_{i},  \label{Me6}
\end{equation}%
\begin{equation}
\mathbf{M}_{4,2}(f_{i}\text{,}\mathbf{v}_{i})=\sum f_{i}\left( \mathbf{v}%
_{i}\cdot \mathbf{v}_{i}+\eta _{i}^{2}\right) \mathbf{v}_{i}\mathbf{v}_{i}%
\text{,}  \label{Me7}
\end{equation}%
\end{subequations}
where $\mathbf{M}_{n}$ means a $n$-th order tensor and $\mathbf{M}_{m,n}$
means a $n$-th-order tensor contracted from a $m$-th order tensor. For the
case of central moments, the variable $\mathbf{v}$ is replaced with $\mathbf{%
v}^{\ast }=\left( \mathbf{v}-\mathbf{u}\right) $. The constraints, (\ref{m1}%
) - (\ref{meq7}), are in fact $16$ linear equations in $f_{i}^{eq}$ in $2$%
-dimensional case and $30$ linear equations in $f_{i}^{eq}$ in $3$%
-dimensional case. Following the same idea as in the definitions, (\ref{Deq1}%
) - (\ref{Deq2}), we further define

\begin{equation}
\boldsymbol{\Delta }_{m,n}\left( \mathbf{v}_{i}\right) =\mathbf{M}%
_{m,n}(f_{i}\text{,}\mathbf{v}_{i})-\mathbf{M}_{m,n}(f_{i}^{eq}\text{,}%
\mathbf{v}_{i})\text{.}  \label{De1}
\end{equation}%
\begin{equation}
\boldsymbol{\Delta }_{m,n}\left( \mathbf{v}_{i}^{\ast }\right) =\mathbf{M}%
_{m,n}(f_{i}\text{,}\mathbf{v}_{i}^{\ast })-\mathbf{M}_{m,n}(f_{i}^{eq}\text{%
,}\mathbf{v}_{i}^{\ast })\text{.}  \label{De2}
\end{equation}%
Except for $\boldsymbol{\Delta }_{0}$, $\boldsymbol{\Delta }_{1}$and $%
\boldsymbol{\Delta }_{2,0}$, the quantity $\boldsymbol{\Delta }_{m,n}$works
as a measure for the deviation of the system from its thermodynamic
equilibrium.

The constraints, (\ref{m1}) - (\ref{meq7}) can also be rewritten as
\begin{equation}
\mathbf{\hat{f}}^{eq}=\mathbf{Cf}^{eq}  \label{meq1}
\end{equation}%
where $\mathbf{\hat{f}}^{eq}=\left[ \hat{f}_{k}^{eq}\right] ^{T}$ and $%
\mathbf{f}^{eq}=\left[ f_{k}^{eq}\right] ^{T}$ are column vectors with $k=1$,%
$2$,$\cdots $,$N$, $\mathbf{C}$ is $N\times N$ matrix whose components are
determined by $\mathbf{v}_{i}$ if the parameter $\eta _{i}$ is fixed. It is
clear that
\begin{equation}
\mathbf{f}^{eq}=\mathbf{C}^{-1}\mathbf{\hat{f}}^{eq}\text{.}  \label{meq2}
\end{equation}%
Obviously, the choosing of the DVM must ensure the existence of $\mathbf{C}%
^{-1}$. The specific choice of the DVM depends on the compromise among the
following several points: (i) numerical efficiency, (ii) numerical
stability, (iii) local symmetry of relevant kinetic moments. We work in the
frame where the particle mass $m=1$ and the constant $R=1$.


If we require the DBM to recover the Navier-Stokes equations in the
continuum limit, the DBM needs a DVM with 3 dimensions.

\subsection{Case with $\protect\gamma =5/3$}

We first consider the simple case where ratio of specific rates is fixed, $%
\gamma =5/3$. We set $\eta _{i}=0$ and $n=0$ in constraint (\ref{meq7}).
Among the seven moment constraints, (\ref{m1}) - (\ref{meq7}), only five are
independent. We do not use the constraints (\ref{m3}) and (\ref{m5}). The
five independent constraints can be rewritten as $26$ independent linear
equations in $f_{i}^{eq}$. Now, we fix the components $\hat{f}_{k}^{eq}$ of $%
\mathbf{\hat{f}}^{eq}$. Here $N=26$.

From the constraint (\ref{m1}), we have $\hat{f}_{1}^{eq}=\rho $. From the
constraint (\ref{m2}), we have $\hat{f}_{2}^{eq}=\rho u_{r}$, $\hat{f}%
_{3}^{eq}=\rho u_{\theta }$, $\hat{f}_{4}^{eq}=\rho u_{\varphi }$. From the
constraint (\ref{m4}), we have $\hat{f}_{5}^{eq}=P+\rho u_{r}^{2}$, $\hat{f}%
_{6}^{eq}=\rho u_{r}u_{\theta }$, $\hat{f}_{7}^{eq}=\rho u_{r}u_{\varphi }$,
$\hat{f}_{8}^{eq}=P+\rho u_{\theta }^{2}$, $\hat{f}_{9}^{eq}=\rho u_{\theta
}u_{\varphi }$, $\hat{f}_{10}^{eq}=P+\rho u_{\varphi }^{2}$. From the
constraint (\ref{m6}), we have $\hat{f}_{11}^{eq}=\rho \left[ T\left(
3u_{r}\right) +u_{r}^{3}\right] $, $\hat{f}_{12}^{eq}=\rho \left( Tu_{\theta
}+u_{r}^{2}u_{\theta }\right) $, $\hat{f}_{13}^{eq}=\rho \left( Tu_{\varphi
}+u_{r}^{2}u_{\varphi }\right) $, $\hat{f}_{14}^{eq}=\rho \left(
Tu_{r}+u_{r}u_{\theta }^{2}\right) $, $\hat{f}_{15}^{eq}=\rho \left(
u_{r}u_{\theta }u_{\varphi }\right) $, $\hat{f}_{16}^{eq}=\rho \left(
Tu_{r}+u_{r}u_{\varphi }^{2}\right) $, $\hat{f}_{17}^{eq}=\rho \left[
T\left( 3u_{\theta }\right) +u_{\theta }^{3}\right] $, $\hat{f}%
_{18}^{eq}=\rho \left[ Tu_{\varphi }+u_{\theta }^{2}u_{\varphi }\right] $, $%
\hat{f}_{19}^{eq}=\rho \left[ Tu_{\theta }+u_{\varphi }^{2}u_{\theta }\right]
$, $\hat{f}_{20}^{eq}=\rho \left[ T\left( 3u_{\varphi }\right) +u_{\varphi
}^{3}\right] $. From the constraint (\ref{meq7}), we have $\hat{f}%
_{21}^{eq}=\rho T\left( 5T+u^{2}\right) +\rho u_{r}^{2}\left(
7T+u^{2}\right) $, $\hat{f}_{22}^{eq}=\rho u_{r}u_{\theta }\left(
7T+u^{2}\right) $, $\hat{f}_{23}^{eq}=\rho u_{r}u_{\varphi }\left(
7T+u^{2}\right) $, $\hat{f}_{24}^{eq}=\rho T\left( 5T+u^{2}\right) +\rho
u_{\theta }^{2}\left( 7T+u^{2}\right) $, $\hat{f}_{25}^{eq}=\rho u_{\theta
}u_{\varphi }\left( 7T+u^{2}\right) $, $\hat{f}_{26}^{eq}=\rho T\left(
5T+u^{2}\right) +2\rho u_{\varphi }^{2}\left( 7T+u^{2}\right) $.

Since the system is spherically symmetric in macroscopic scale, $u_{\theta
}=u_{\varphi }=0$ and $u^{2}=u_{r}^{2}$ in the above expressions for $%
\mathbf{\hat{f}}^{eq}=\left[ \hat{f}_{1}^{eq}\text{, }\hat{f}_{2}^{eq}\text{,%
}\cdots \text{,}\hat{f}_{N}^{eq}\right] ^{T}$.

The components of the matrix $\mathbf{C=}\left[ \mathbf{C}_{k}\right]
\mathbf{=}\left[ C_{ki}\right] $ should be fixed in the same sequence, where
$k=1$,$2$,$\cdots $,$26$ and $i=1$,$2$,$\cdots $,$26$. From the constraint (%
\ref{m1}),we have $C_{1i}=1$. From the constraint (\ref{m2}), we have $%
C_{2i}=v_{ir}$, $C_{3i}=v_{i\theta }$, $C_{4i}=v_{i\varphi }$. From the
constraint (\ref{m4}), we have $C_{5i}=v_{ir}^{2}$, $C_{6i}=v_{ir}v_{i\theta
}$, $C_{7i}=v_{ir}v_{i\varphi }$, $C_{8i}=v_{i\theta }^{2}$, $%
C_{9i}=v_{i\theta }v_{i\varphi }$, $C_{10i}=v_{i\varphi }^{2}$. From the
constraint (\ref{m6}), we have $C_{11i}=v_{ir}^{3}$, $C_{12i}=v_{ir}^{2}v_{i%
\theta }$, $C_{13i}=v_{ir}^{2}v_{i\varphi }$, $C_{14i}=v_{ir}v_{i\theta
}^{2} $, $C_{15i}=v_{ir}v_{i\theta }v_{i\varphi }$, $C_{16i}=v_{ir}v_{i%
\varphi }^{2}$, $C_{17i}=v_{i\theta }^{3}$, $C_{18i}=v_{i\theta
}^{2}v_{i\varphi }$, $C_{19i}=v_{i\theta }v_{i\varphi }^{2}$, $%
C_{20i}=v_{i\varphi }^{3}$. From the constraint (\ref{meq7}), we have $%
C_{21i}=\left( v_{ir}^{2}+v_{i\theta }^{2}+v_{i\varphi }^{2}\right)
v_{ir}^{2}$, $C_{22i}=\left( v_{ir}^{2}+v_{i\theta }^{2}+v_{i\varphi
}^{2}\right) v_{ir}v_{i\theta }$, $C_{23i}=\left( v_{ir}^{2}+v_{i\theta
}^{2}+v_{i\varphi }^{2}\right) v_{ir}v_{i\varphi }$, $C_{24i}=\left(
v_{ir}^{2}+v_{i\theta }^{2}+v_{i\varphi }^{2}\right) v_{i\theta }^{2}$, $%
C_{25i}=\left( v_{ir}^{2}+v_{i\theta }^{2}+v_{i\varphi }^{2}\right)
v_{i\theta }v_{i\varphi }$, $C_{26i}=\left( v_{ir}^{2}+v_{i\theta
}^{2}+v_{i\varphi }^{2}\right) v_{i\varphi }^{2}$.

\begin{figure}
  \centering
  \includegraphics[width=0.8\textwidth]{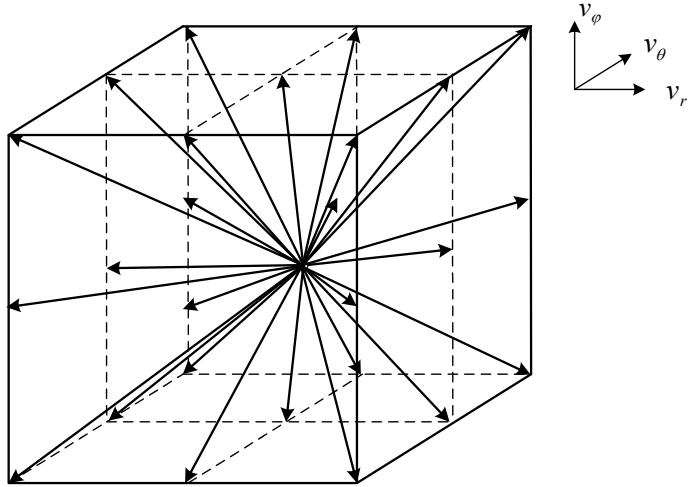}\\
  \caption{Schematic of the discrete velocity model (D3V26).}\label{fig0-2}
\end{figure}

An example for the 3-Dimensional 26-Velocity(D3V26) DVM is as below,
\begin{subequations}
\begin{eqnarray}
\mathbf{v}_{i} &=&\left\{
\begin{array}{cc}
\left( 0,\pm 1,\pm 1\right) c_{1} & i=1,\cdots ,4 \\
\left( \pm 1,0,\pm 1\right) c_{1} & i=5,\cdots ,8 \\
\left( \pm 1,\pm 1,0\right) c_{1} & i=9,\cdots ,12%
\end{array}%
\right. \\
\mathbf{v}_{i} &=&\left\{
\begin{array}{cc}
\left( \pm 1,\pm 1,\pm 1\right) c_{2} & i=13,\cdots ,20%
\end{array}%
\right. \\
\mathbf{v}_{i} &=&\left\{
\begin{array}{cc}
\left( \pm 1,0,0\right) c_{3} & i=21,22 \\
\left( 0,\pm 1,0\right) c_{3} & i=23,24 \\
\left( 0,0,\pm 1\right) c_{3} & i=25,26%
\end{array}%
\right. .
\end{eqnarray}
\end{subequations}
The schematic of the discrete velocity model is shown in Figure \ref{fig0-2}.
The expressions for the inverse of the
matrix $\mathbf{C}$, $\mathbf{C}^{-1}=\left[ \mathbf{C}_{k}^{-1}\right] $,
can be easily obtained by using the software, MatLab, then a specific example of the DVM is formulated.

\begin{figure}
  \centering
  \includegraphics[width=0.6\textwidth]{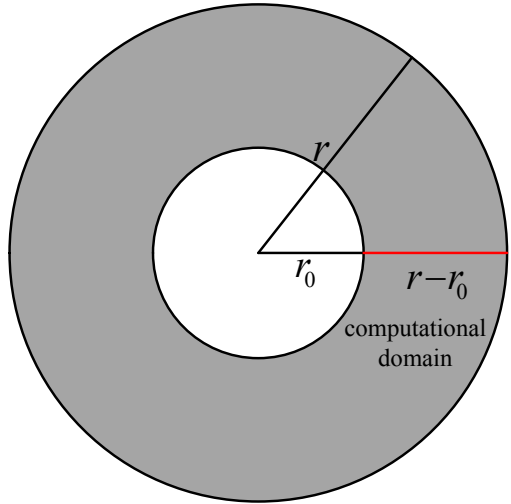}\\
  \caption{Schematic of the computational domain (gray area).}\label{fig0-3}
\end{figure}

Up to this step, a specific discretization
of the velocity space has been performed. Consequently, the DBM for system
with spherical symmetry and $\gamma =5/3$ has been constructed. The spatial and temporal derivatives of the distribution function in the kinetic model can be calculated in the normal way. If we are not interested in the extra degrees of freedom other than the translational, the formulated DBM can be used to study the hydrodynamic and the thermodynamic behaviors of the compressible flow system. The computational domain in this work can be found in Figure \ref{fig0-3} where the projection of computational domain in two-dimensional space is shown. In the rest of the article, $r-r_0$ is used as the label of the space axis where $r_0$ is the distance between the computational domain and the centre of sphere.

Firstly, we simulate a Sod shock tube problem using the DBM where the ``force term'' in Eq. (\ref{DBS1}) does not exist. Such a test can be used to check the validity of the DVM. In addition, the case with  ``force term'' vanished corresponds the case where the value of $r_0$ is so large that the geometric effects are negligible.
\begin{figure}
  \centering
  \includegraphics[width=0.8\textwidth]{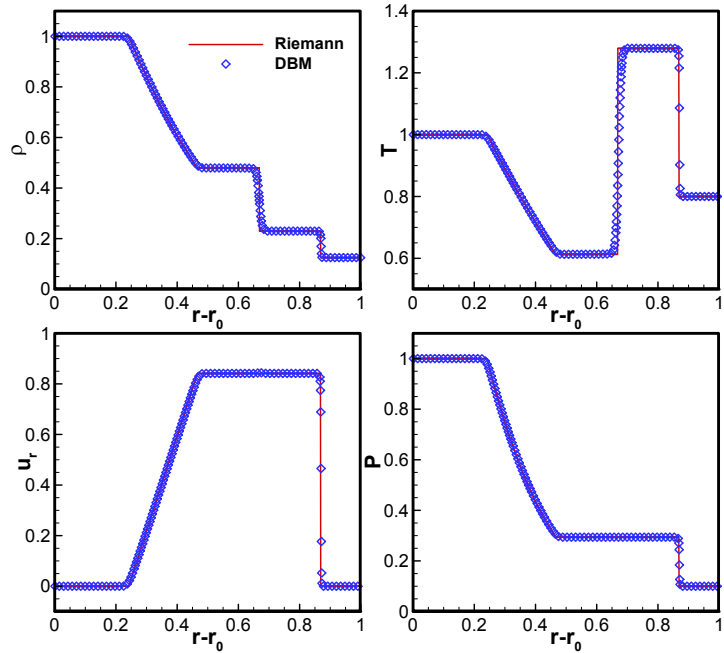}\\
  \caption{Profile of macroquantities for Sob shock tube with $\gamma = 5/3$.}\label{fig1}
\end{figure}
The results are shown in Figure \ref{fig1}. The Riemann solutions are also plotted for comparison. The calculated results show that the new DVM (D3V26) is applicable and the new DBM model can capture the discontinuous interface.

\begin{figure}
  \centering
  \includegraphics[width=0.9\textwidth]{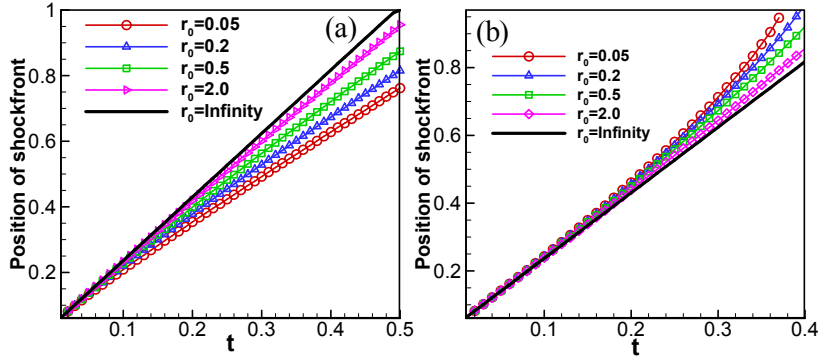}\\
  \caption{The position of shock wavefront versus time for the case with $\gamma =5/3$. (a)Shock wave propagates outward (explosion).(b)Shock wave propagates inward (implosion).}\label{fig2}
\end{figure}

\begin{figure}
  \centering
  \includegraphics[width=0.9\textwidth]{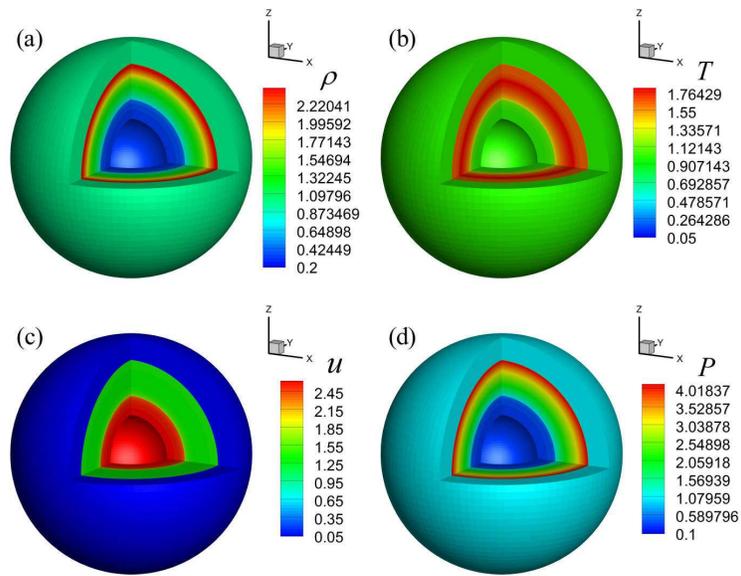}\\
  \caption{ The contour maps of macroquantities for shock wave propagating outward  for $r_0=0.5$ at $t=0.9$. (a)Contour map of density, $\rho$. (b)Contour map of temperature, $T$. (c)Contour map of velocity, $u$. (d)Contour map of pressure, $P$.}\label{fig2-1}
\end{figure}

Then, to check the geometric effects, two cases of propagating shock wave with Mach number $Ma=1.5$ along the radial direction are simulated. In the first case the shock wave propagates outward so that the divergence makes effects. Several cases with different values of $r_0$ are simulated. The results are shown in Figure \ref{fig2}(a). For easy understanding, the three dimensional contour maps of hydrodynamic quantities at a certain time for the case with $r_0=0.5$ are given in Figure \ref{fig2-1}. Figure \ref{fig2}(a) shows the variation of position of wavefront with time. From the figure, we can find that the propagation speed of shock wave decreases with time due to the divergence effect. Besides, the smaller $r_0$ has a faster decrease of propagation speed. The second case is that the shock wave propagates inward so the convergence effect plays a role. Figure \ref{fig2}(b) shows the variation of position of wavefront with time. The propagation speed of shock wave increases with  time due to the convergence effect, and the smaller $r_0$ has a faster increase of propagation speed.

The above simulation results and analysis show that the new model can well describe the geometric effects of the flows with spherical symmetry.

\subsection{Case with flexible $\protect\gamma $}

For the case with flexible ratio of specific heat $\gamma $, if we are interested only in the hydrodynamic behaviors, we can use the simulation results of the DBM
formulated in last subsection. Just get the number $n$ of the extra degree
of freedom using its relation to $\gamma $, then obtain the total internal
energy $E$ using its definition. If we are interested also in the
thermodynamic nonequilibrium behavior, we need to continue the formulation of DBM.

To model the case with flexible $\gamma $, we resort to the parameter $\boldsymbol{\eta}$ to describe the contribution of extra degrees of freedom. The distribution function $f$ with $n$ extra degrees of freedom  can be replaced by $g$ and $h$, and the equilibrium distribution function $f^{eq}$ in Eq.(\ref{feq}) is replaced by $g^{eq}$ and $h^{eq}$, where
\begin{subequations}
\begin{eqnarray}
  g &=& \int fd\boldsymbol{\eta}, \\
  h &=& \int f \eta ^2d\boldsymbol{\eta}, \\
  g^{eq} &=& \int f^{eq}d\boldsymbol{\eta}, \\
  h^{eq} &=& \int f^{eq} \eta ^2d\boldsymbol{\eta}.   \label{eqgh}
\end{eqnarray}
\end{subequations}
 The functions $g$ and $g^{eq}$ recover to the $f$ and $f^{eq}$ when there is no extra degrees of freedom (i.e., $\gamma =5/3$).
The evolution equation of $f$ in Eq.(\ref{eq22}) becomes two evolution equations of $g$ and $h$. They  read
\begin{subequations}
\begin{eqnarray}
 \partial _{t}g+v_{r}\frac{\partial g}{\partial r}+\left[ \frac{%
v_{r}v_{\theta }^{2}}{rT}+\frac{v_{r}v_{\varphi }^{2}}{rT}-\frac{\left(
v_{\theta }^{2}+v_{\varphi }^{2}\right) \left( v_{r}-u_{r}\right) }{rT}%
\right] g^{eq}=-\frac{1}{\tau }\left( g-g^{eq}\right) ,  \label{gevolu}\\
  \partial _{t}h+v_{r}\frac{\partial h}{\partial r}+\left[ \frac{%
v_{r}v_{\theta }^{2}}{rT}+\frac{v_{r}v_{\varphi }^{2}}{rT}-\frac{\left(
v_{\theta }^{2}+v_{\varphi }^{2}\right) \left( v_{r}-u_{r}\right) }{rT}%
\right] h^{eq}=-\frac{1}{\tau }\left( h-h^{eq}\right) .  \label{hevolu}
\end{eqnarray}
\end{subequations}
where
\begin{subequations}
\begin{eqnarray}
g^{eq}\left( \mathbf{v}\right) =\rho \left( \frac{1}{2\pi RT}\right)
^{D/2}\exp \left[ -\frac{\left( \mathbf{v}-\mathbf{u}\right) ^{2}}{2RT}\right] \text{,}\\
h^{eq}= \frac{nT}{2}g^{eq} . {\kern 58pt}  \label{geq}
\end{eqnarray}
\end{subequations}
As a result, the $f_i^{eq}$ solved before, for the case with $\gamma =5/3$, can be used as $g_i^{eq}$ here. The $h_i^{eq}$ can be solved by
\begin{equation}
h_i^{eq}= \frac{nT}{2}g_i^{eq}    \text{.}  \label{eqhieq}
\end{equation}%
In this case, the discretization of $\boldsymbol{\eta}$ does not need at all, and the moments in Eq.(\ref{Me1})-(\ref{Me7}) are calculated by
\begin{subequations}
\begin{equation}
\mathbf{M}_{0}(f_{i}\text{,}\mathbf{v}_{i})=\sum_{i=1}^{N}g_{i}\text{,}
\label{eqghMe1}
\end{equation}%
\begin{equation}
\mathbf{M}_{1}(f_{i}\text{,}\mathbf{v}_{i})=\sum_{i=1}^{N}g_{i}\mathbf{v}_{i}%
\text{,}  \label{eqghMe2}
\end{equation}%
\begin{equation}
\mathbf{M}_{2,0}(f_{i}\text{,}\mathbf{v}_{i})=\sum_{i=1}^{N}[g_{i}(\mathbf{v}%
_{i}\cdot \mathbf{v}_{i})+h_i]\text{,}  \label{eqghMe3}
\end{equation}%
\begin{equation}
\mathbf{M}_{2}(f_{i}\text{,}\mathbf{v}_{i})=\sum_{i=1}^{N}g_{i}\mathbf{v}_{i}%
\mathbf{v}_{i}\text{,}  \label{eqghMe4}
\end{equation}%
\begin{equation}
\mathbf{M}_{3,1}(f_{i}\text{,}\mathbf{v}_{i})=\sum_{i=1}^{N}[g_{i}(\mathbf{v}%
_{i}\cdot \mathbf{v}_{i})+h_i] \mathbf{v}_{i}\text{,}  \label{eqthMe5}
\end{equation}%
\begin{equation}
\mathbf{M}_{3}(f_{i}\text{,}\mathbf{v}_{i})=\sum g_{i}\mathbf{v}_{i}\mathbf{v%
}_{i}\mathbf{v}_{i},  \label{eqghMe6}
\end{equation}%
\begin{equation}
\mathbf{M}_{4,2}(f_{i}\text{,}\mathbf{v}_{i})=\sum \left[g_{i}\left( \mathbf{v}%
_{i}\cdot \mathbf{v}_{i}\right)+h_i \right] \mathbf{v}_{i}\mathbf{v}_{i}%
\text{,}  \label{eqghMe7}
\end{equation}%
\end{subequations}
The constraints, (\ref{eqghMe1})-(\ref{eqghMe7}), are 16 linear equations in 2-dimensional case and 30 linear equations equations in 3-dimensional case. The definitions of $\mathbf{\Delta} _{m,n}(\mathbf{v}_i)$ and $\mathbf{\Delta} _{m,n}(\mathbf{v}^*_i)$ in Eq.(\ref{De1}) and (\ref{De2}) still keep the same and work as a measure for the
deviation of the system from its thermodynamic equilibrium. Up to this step, a DBM with
D3V26 for system with flexible specific heat ratio $\gamma $ has been formulated.

\begin{figure}
  \centering
  \includegraphics[width=0.8\textwidth]{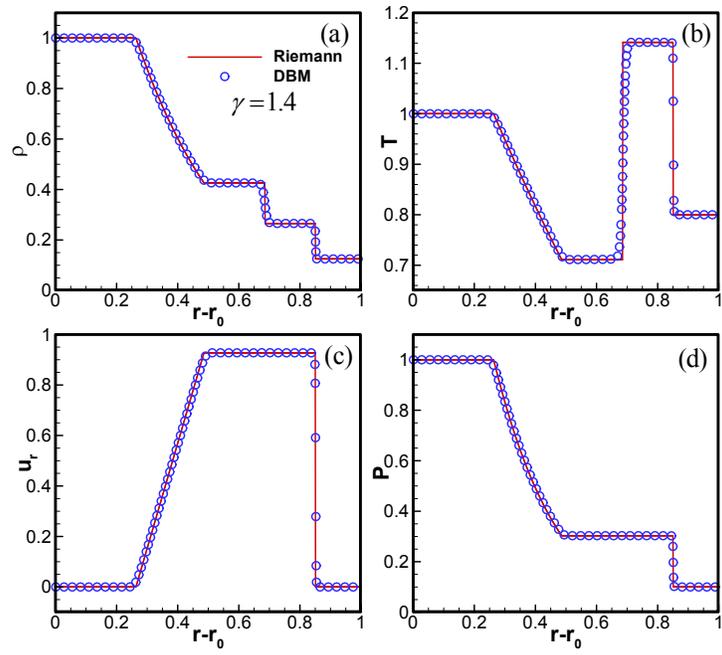}\\
  \caption{Profiles of macroquantities for Sob shock tube with $\gamma = 1.4$ ($n=2$).}\label{fig3}
\end{figure}

\begin{figure}
  \centering
  \includegraphics[width=0.8\textwidth]{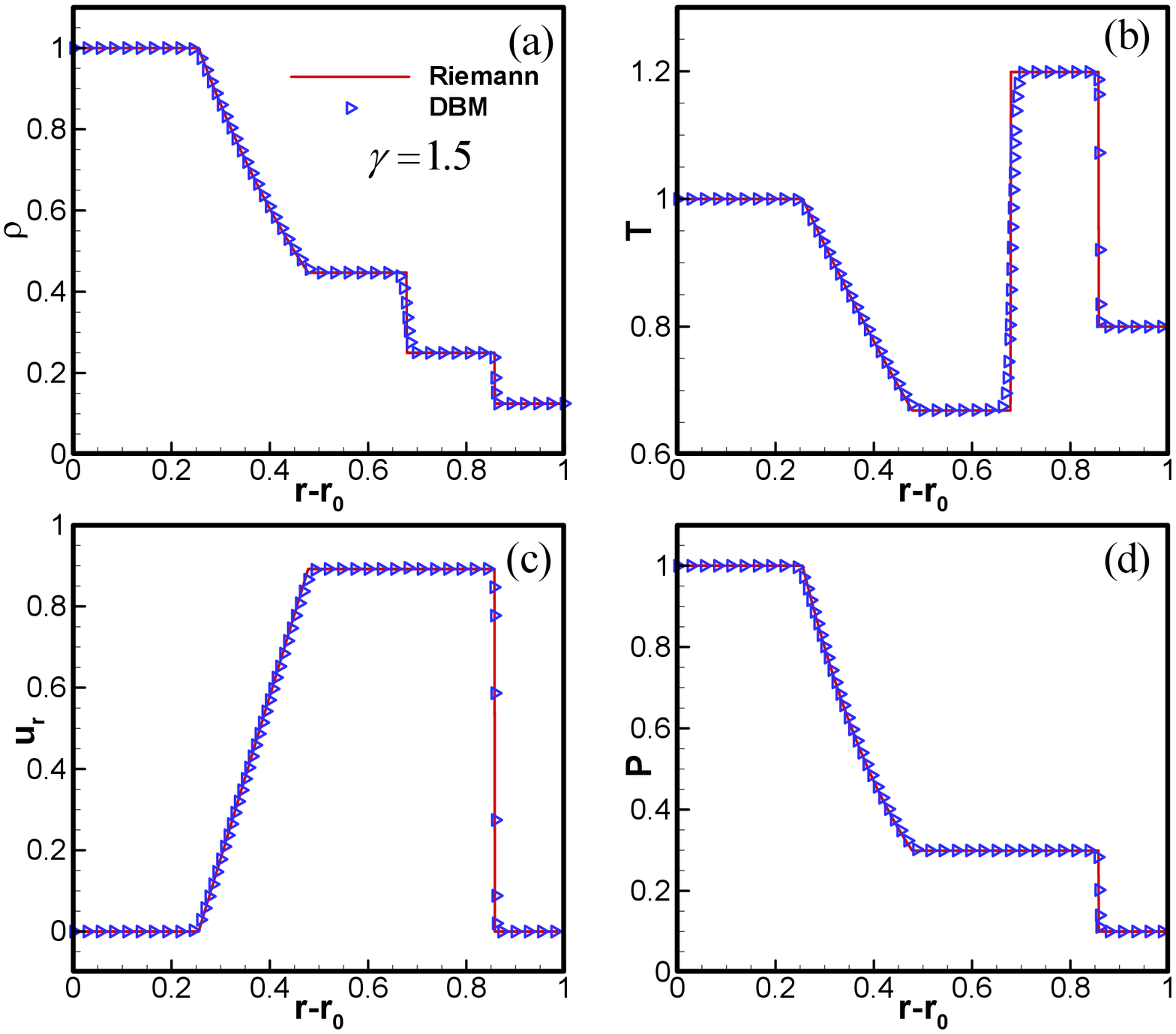}\\
  \caption{Profiles of macroquantities for Sob shock tube with $\gamma = 1.5$ ($n=1$).}\label{fig3-2}
\end{figure}

In order to test the flexibility of $\gamma$, Sod shock tube problems with two different values of $\gamma$ are simulated. The specific heat ratio for one case is $\gamma=1.4$ which means there are two extra degrees of freedom ($n=2$), and the other case is $\gamma=1.5$ which means there is only one extra degree of freedom ($n=1$). First ignore the geometric effects. The results for the two cases are shown Figure \ref{fig3} and Figure \ref{fig3-2}, respectively. The results are in well agreement with Riemann solutions which verifies the validity of the new model.

\begin{figure}
  \centering
  \includegraphics[width=0.9\textwidth]{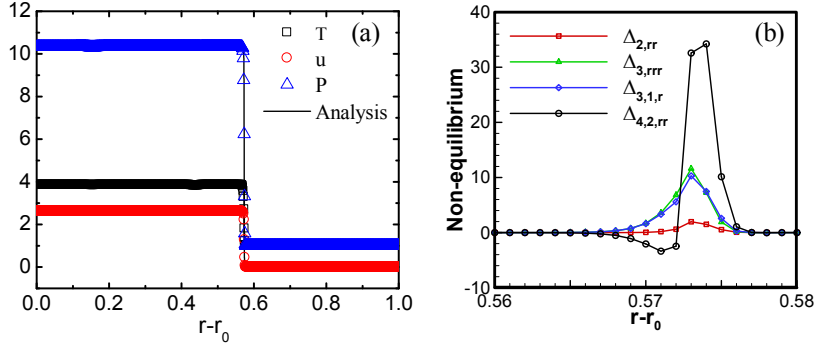}\\
  \caption{Profiles of macroquantities and non-equilibrium effects around shock wavefront for the case with $Ma=3.0$ and $\gamma=1.4$.(a)Profiles of $\rho$, $u$, and $P$. The symbols indicate the results of DBM and solid lines are solutions based on Hugoniot relations of shock wave.(b)Profiles of non-equilibrium effects around shock wavefront.}\label{fig4}
\end{figure}

Then, a shock wave with $Ma=3$ propagating along the radial direction is simulated for the case with  $\gamma=1.4$ (i.e., $n=2$). Firstly, we assume that shock wave starts at infinity from the centre of sphere, so the ``force term'' in Eqs.(\ref{gevolu}) and (\ref{hevolu}) are negligible. The profiles of the macroquantities, including $\rho$, $u$, and $P$ around wavefront at a certain time are shown in Figure \ref{fig4}(a). The macroquantities on the two sides of wavefront satisfy the Hugoniot relations of shock wave. The corresponding non-equilibrium quantities including $\Delta_{2,rr}(v_i)$, $\Delta_{3,rrr}(v_i)$, $\Delta_{3,1,r}(v_i)$, and $\Delta_{4,2,rr}(v_i)$ around the wavefront are plotted in Figure \ref{fig4}(b). Since the shock propagate along the radial direction, only one component in $r$ direction is considered for each of the four kinds non-equilibrium quantities. From Figure \ref{fig4}(b), we can get the non-equilibrium characteristics from a different point of view, other than the viscous stress and heat flux, which can not be provided by the traditional Navier-Stokes equations.

\begin{figure}
  \centering
  \includegraphics[width=0.9\textwidth]{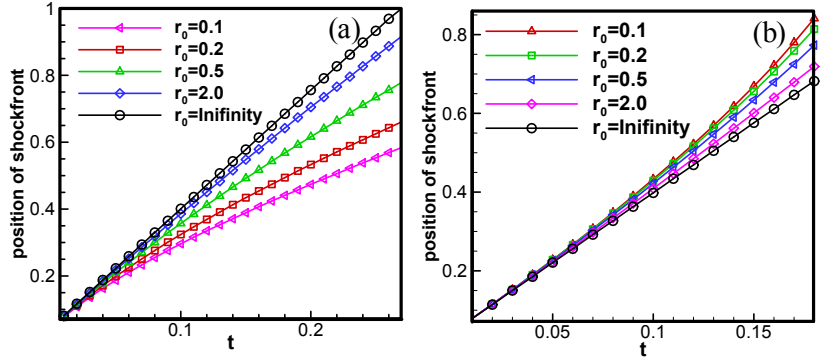}\\
  \caption{The position of shock wavefront versus time for the case with $\gamma=1.4$. (a)Shock wave propagates outward.(b)Shock wave propagates inward.}\label{fig5}
\end{figure}

Then the geometric effects are taken into account. Shock waves propagating outward and inward are both simulated. The evolutions of positions of wavefront are shown in Figures \ref{fig5}(a) and \ref{fig5}(b), respectively.

From Figure \ref{fig5}(a), it can be seen the shock wave has a constant propagation speed for $r_0=\mathrm{Infinity}$ which corresponds to the case in Figure \ref{fig4}. For the shock wave propagating outward, the propagation speed decrease with time, and the smaller $r_0$ has a faster decrease which indicates a stronger ``force'' caused by geometric effects. The shock wave propagating inward has a similar characteristics, except that the propagation speed increase with time due to the  convergence effect.

\subsection{Thermodynamic non-equilibrium characteristics in explosion and implosion}

\begin{figure}
  \centering
  \includegraphics[width=0.9\textwidth]{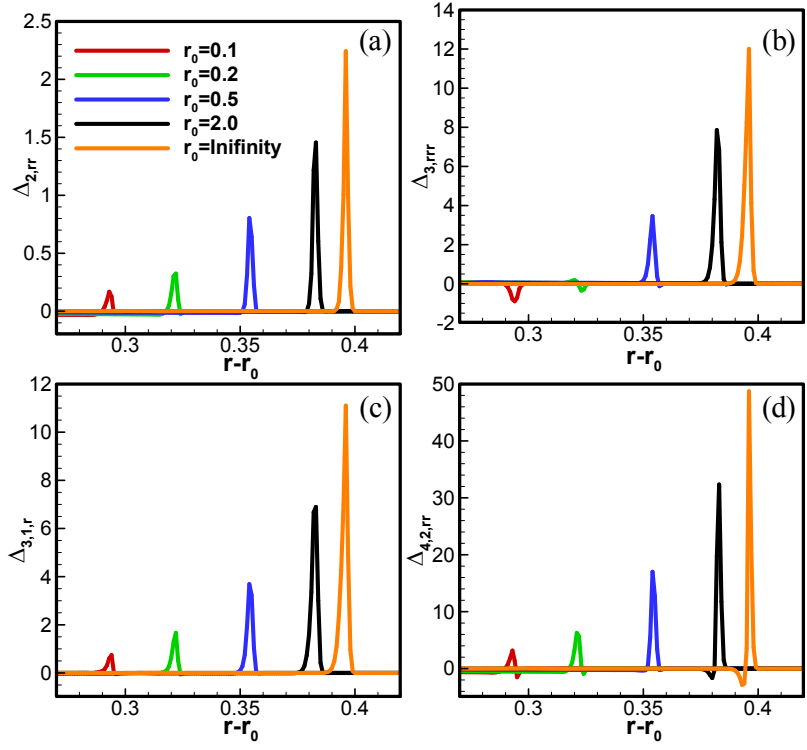}\\
  \caption{The non-equilibrium quantities around wavefront for explosion.}\label{fig6}
\end{figure}

The non-equilibrium quantities around wavefront for the shock propagating outward (i.e., explosion), at a certain time, are plotted in Figure \ref{fig6}. In this case, the divergence effect plays the role of resistance force. It decreases the propagation speed and reduce the non-equilibrium strength. It can be found in Figure \ref{fig6}(b) that a smaller $r_0$, which means a stronger divergence effect, may even change the direction of deviation from equilibrium state for $\Delta_{3,rrr}$.

It is interesting to have more discussions on the reasonability of the
approximation $f=f^{eq}$ in treating with the term for pure geometric effects in Eq.(\ref{eq19}). When the flow under consideration is not close to the spherical center, the error induced by the approximation when calculating the ``force term" is relatively very small compared with the ``convective term``.
 The comparison can be made in the moment space.

\begin{figure}
  \centering
  \includegraphics[width=1.0\textwidth]{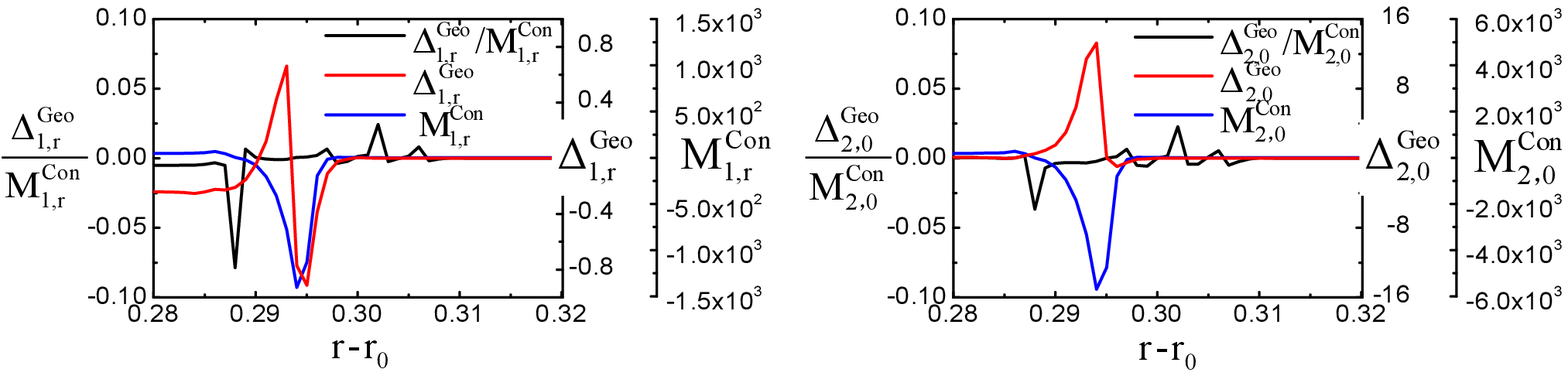}\\
  \caption{The profiles of (a)$\Delta _{1,r}^{Geo}/M_{1,r}^{Con}$, $\Delta _{1,r}^{Geo}$, and $M_{1,r}^{Con}$  and  (b)$\Delta _{2,0}^{Geo}/M_{2,0}^{Con}$, $\Delta _{2,0}^{Geo}$, and $M_{2,0}^{Con}$ around the shock wave front for the case with ${r_0} = 0.1$ (explosion).}\label{explosion-fig-add1}
\end{figure}

\begin{figure}
  \centering
  \includegraphics[width=1.0\textwidth]{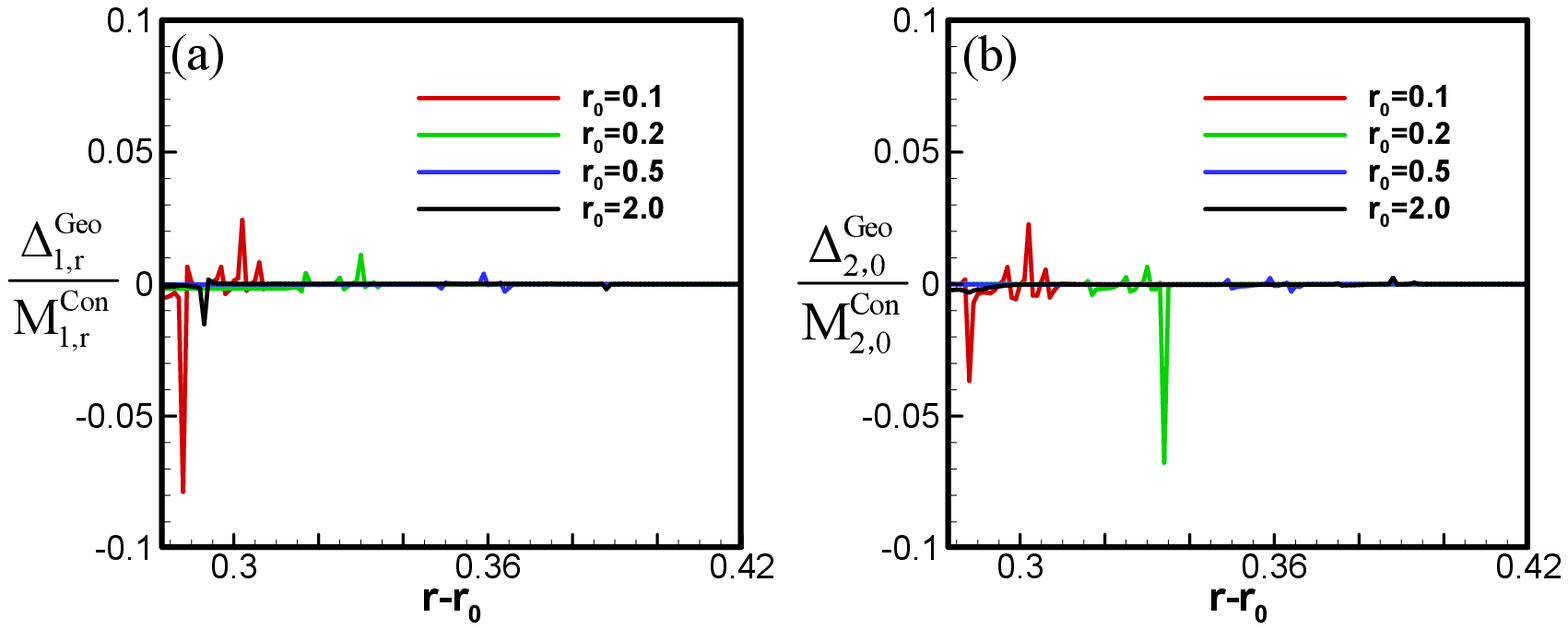}\\
  \caption{The profiles of (a)$\Delta _{1,r}^{Geo}/M_{1,r}^{Con}$ and (b)$\Delta _{2,0}^{Geo}/M_{2,0}^{Con}$ around the shock wave front for different cases (explosion). }\label{explosion-fig-add2}
\end{figure}

Firstly, the ``force term" in moment space is defined as
\begin{equation}\label{Eq-MnGeo}
 \mathbf{M}_n^{Geo}(f) = \int {\left( {\frac{{v_\theta ^2 + v_\varphi ^2}}{r}\frac{{\partial f}}{{\partial {v_r}}} - \frac{{{v_r}{v_\theta }}}{r}\frac{{\partial f}}{{\partial {v_\theta }}} - \frac{{{v_r}{v_\varphi }}}{r}\frac{{\partial f}}{{\partial {v_\varphi }}}} \right){{\bf{v}}^n}d{\bf{v}}},
\end{equation}
where the subscript $n$ in $\mathbf{M}_n^{Geo}(f)$ indicates the $n$-th order moment. Then the error induced by the approximation $f=f^{eq}$ in the moment space is
\begin{equation}\label{Eq-DeltanGeo}
\pmb{\Delta} _n^{Geo} = \mathbf{M}_n^{Geo}(f) - \mathbf{M}_n^{Geo}({f^{eq}}),
\end{equation}
Similarly, the convective term in the moment space is defined as
\begin{equation}\label{Eq-MnCon}
\mathbf{M}_n^{Con}(f) = \int {\left( {{\bf{v}} \cdot \nabla f} \right){{\bf{v}}^n}d{\bf{v}}}.
\end{equation}
The ratio $\Delta _n^{Geo}/M_n^{Con}(f)$ measures the relative error induced by $f=f^{eq}$ when calculating the ``force term" with respect to the convective term. Since the density, momentum, and energy is the 0-th order, 1st order, and 2nd order moments, respectively, only the $\Delta _0^{Geo}/M_0^{Con}$, $\Delta _{1,r}^{Geo}/M_{1,r}^{Con}$, and $\Delta _{2,0}^{Geo}/M_{2,0}^{Con}$ are investigated here. The subscript ``2,0" indicates that $\Delta _{2,0}^{Geo}$ (or $M_{2,0}^{Con}$) is a 0-th order tensor (i.e., scalar) contracted from the 2nd order tensor. According to the definition of $\pmb{\Delta} _n^{Geo}$ and $\mathbf{M}_n^{Con}(f)$, we can get that
\begin{equation}\label{Eq-DetlaM0}
\frac{{\Delta _0^{Geo}}}{{M_0^{Con}}} = \frac{{\int {2\frac{{{v_r}}}{r}\left( {f - {f^{eq}}} \right)d{\bf{v}}} }}{{\int {\left( {{\bf{v}} \cdot \nabla f} \right)d{\bf{v}}} }},
\end{equation}

\begin{equation}\label{Eq-DetlaM1}
\frac{{\Delta _{1,r}^{Geo}}}{{M_{1,r}^{Con}}} = \frac{{\int {\left( { - \frac{{v_\theta ^2 + v_\varphi ^2}}{r} + 2\frac{{{v_r}{v_\theta }}}{r} + 2\frac{{{v_r}{v_\varphi }}}{r}} \right)(f - {f^{eq}})d{\bf{v}}} }}{{\frac{{\partial {M_{2,rr}}}}{{\partial r}} + \frac{{\partial {M_{2,r\theta }}}}{{\partial \theta }} + \frac{{\partial {M_{2,r\varphi }}}}{{\partial \varphi }}}},
\end{equation}

\begin{equation}\label{Eq-DetlaM20}
\frac{{\Delta _{2,0}^{Geo}}}{{M_{2,0}^{Con}}} = \frac{{\int {\left[ {\left( { - 2{v_r}\frac{{v_\theta ^2 + v_\varphi ^2}}{r}} \right) + \left( {\frac{{{v_r}^3 + 3{v_r}v_\theta ^2 + {v_r}v_\varphi ^2}}{r}} \right) + \left( {\frac{{{v_r}^3 + {v_r}v_\theta ^2 + 3{v_r}v_\varphi ^2}}{r}} \right)} \right](f - {f^{eq}})d{\bf{v}}} }}{{\frac{{\partial {M_{3.1,r}}}}{{\partial r}} + \frac{{\partial {M_{3.1,\theta }}}}{{\partial \theta }} + \frac{{\partial {M_{3.1,\varphi }}}}{{\partial \varphi }}}}.
\end{equation}
It is obvious that $\Delta _0^{Geo}/M_0^{Con}$ is always equal to zero. Therefore, only the $\Delta _{1,r}^{Geo}/M_{1,r}^{Con}$ and $\Delta _{2,0}^{Geo}/M_{2,0}^{Con}$ are used to measure the relative error of ''force term``. The profiles of $\Delta _{1,r}^{Geo}/M_{1,r}^{Con}$, $\Delta _{1,r}^{Geo}$, and $M_{1,r}^{Con}$ for the case with $r_0=0.1$ is shown in Figure \ref{explosion-fig-add1}(a). It can be seen that $\Delta _{1,r}^{Geo}$ is so small compared with $M_{1,r}^{Con}$ that it nearly can be ignored even for such a small $r_0$. From the Figure \ref{explosion-fig-add1}(b) we can see the $\Delta _{2,0}^{Geo}$  is also nearly negligible compared with $M_{2,0}^{Con}$. Besides, with the increase of $r_0$, the effects become weaker, the $\Delta _n^{Geo}$ is more negligible compared with $M_n^{Con}$, which is verified in Figure \ref{explosion-fig-add2}. From Figure \ref{explosion-fig-add2} we can conclude that the error $\Delta _n^{Geo}$ is negligible compared with $M_n^{Con}$ when $r_0 \geq 0.1$.

\begin{figure}
  \centering
  \includegraphics[width=0.9\textwidth]{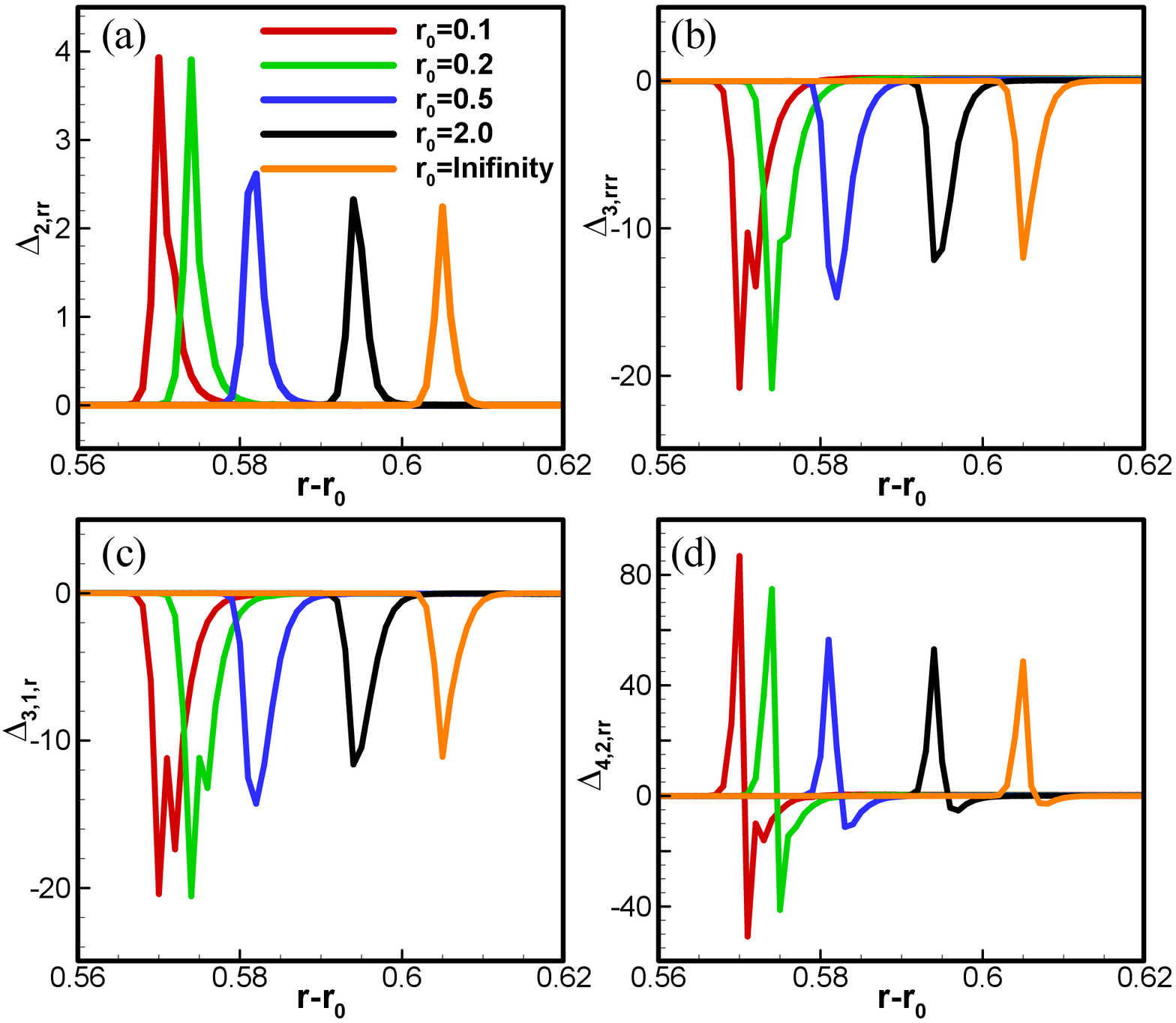}\\
  \caption{The non-equilibrium quantities around wavefront for implosion.}\label{fig7}
\end{figure}

For the shock propagating inward, the non-equilibrium quantities at a certain time are shown in Figure \ref{fig7}. In this case, the convergence effect accelerates the propagation speed of the shock wave. So the non-equilibrium effects around wavefront are also strengthened. A smaller $r_0$ corresponds to a stronger non-equilibrium. Unlike the shock propagating outward, no change of the direction for the deviation from thermodynamic equilibrium state is found.

\begin{figure}
  \centering
  \includegraphics[width=1.0\textwidth]{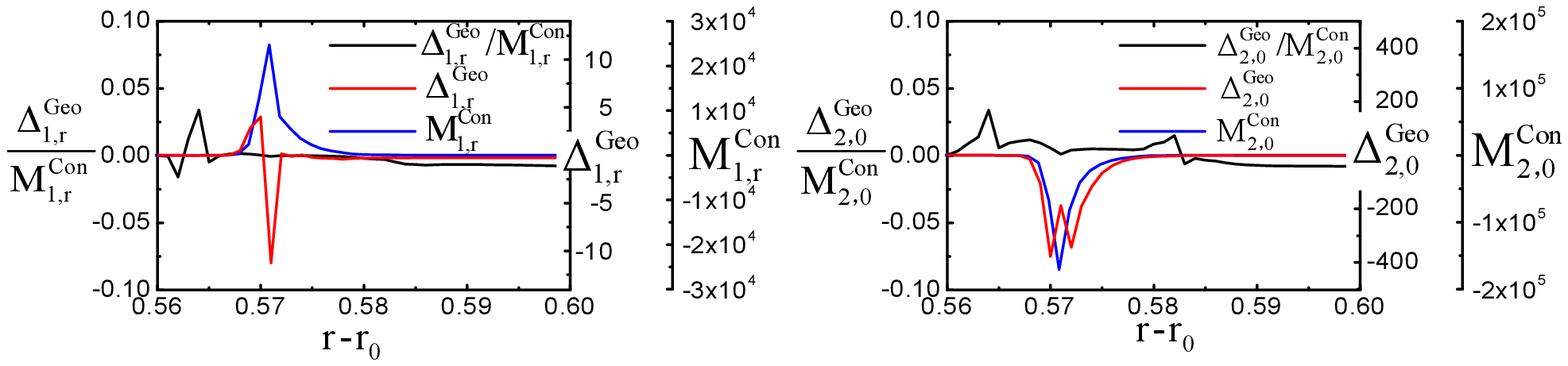}\\
  \caption{The profiles of (a)$\Delta _{1,r}^{Geo}/M_{1,r}^{Con}$, $\Delta _{1,r}^{Geo}$, and $M_{1,r}^{Con}$  and  (b)$\Delta _{2,0}^{Geo}/M_{2,0}^{Con}$, $\Delta _{2,0}^{Geo}$, and $M_{2,0}^{Con}$ around the shock wave front for the case with ${r_0} = 0.1$ (implosion).}\label{implosion-fig-add1}
\end{figure}

\begin{figure}
  \centering
  \includegraphics[width=1.0\textwidth]{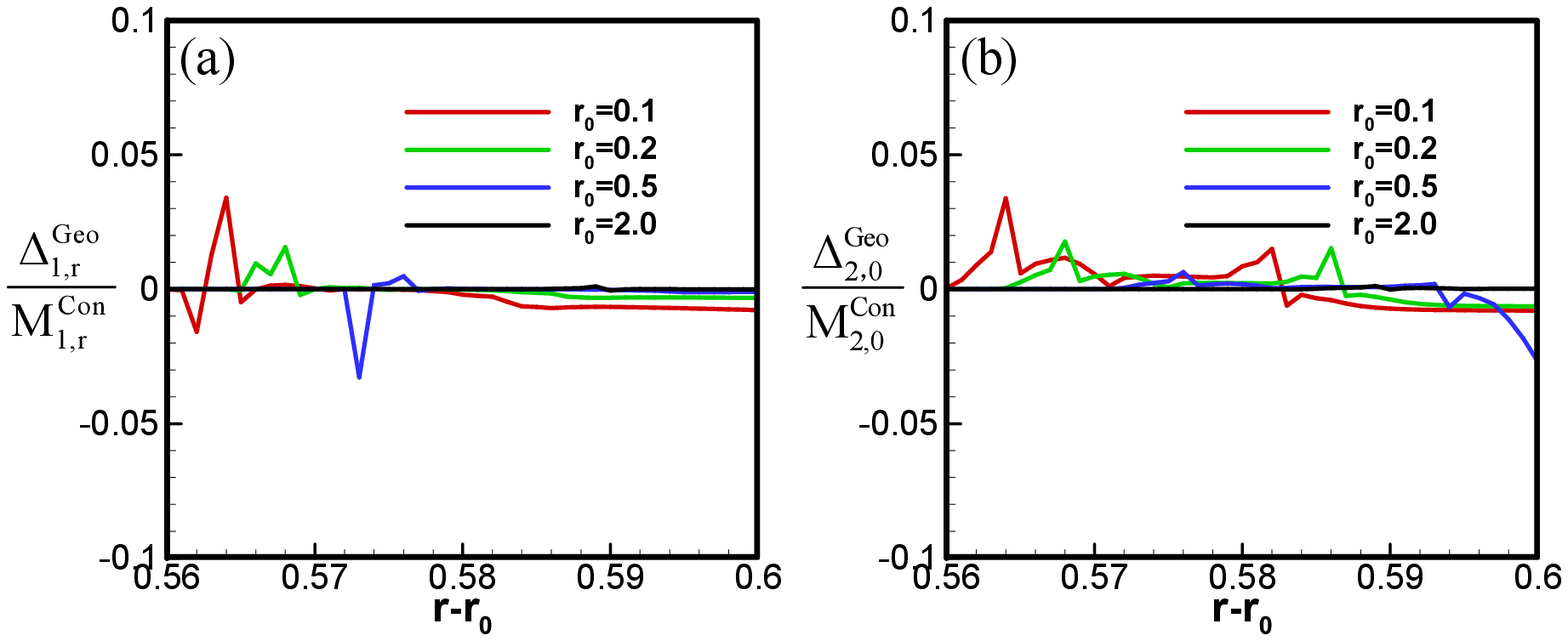}\\
  \caption{The profiles of (a)$\Delta _{1,r}^{Geo}/M_{1,r}^{Con}$ and (b)$\Delta _{2,0}^{Geo}/M_{2,0}^{Con}$ around the shock wave front for different cases (implosion).}\label{implosion-fig-add2}
\end{figure}

The comparisons between the error caused by the approximation $f=f^{eq}$ when calculating the ``force term" and the convective term are given in Figure \ref{implosion-fig-add1} and Figure \ref{implosion-fig-add2}. The profiles of $\Delta _{1,r}^{Geo}/M_{1,r}^{Con}$, $\Delta _{1,r}^{Geo}$, and $M_{1,r}^{Con}$ for the case with $r_0=0.1$ is shown in Figure \ref{implosion-fig-add1}(a). It can be seen that $\Delta _{1,r}^{Geo}$ is so small compared with $M_{1,r}^{Con}$ that it nearly can be ignored even for such a small $r_0$. From the Figure \ref{implosion-fig-add1}(b) we can see the $\Delta _{2,0}^{Geo}$  is also nearly negligible compared with $M_{2,0}^{Con}$. Besides, with the increase of $r_0$, the effects become weaker, the $\Delta _n^{Geo}$ is more negligible compared with $M_n^{Con}$, which is verified in Figure \ref{implosion-fig-add2}. From Figure \ref{implosion-fig-add2} we can conclude that the error $\Delta _n^{Geo}$ is negligible compared with $M_n^{Con}$ when $r_0 \geq 0.1$ for the current implosion numerical experiment setup.

\section{Conclusion and discussions}

We present a theoretical framework for constructing DBM
in spherical coordinates for the compressible flow systems
with spherical symmetry. A key technique here is to use \emph{local} Cartesian
coordinates to describe the particle velocity. Thus, compared with the Boltzmann equation in Cartesian coordinates, the geometric effects, like the divergence and
convergence, are treated as a ``force
term''.
For such a system, even though the hydrodynamic model is one-dimensional, the DBM needs a discrete velocity model
with 3 dimensions. A new scheme is introduced so that the DBM can use the same set of discrete velocities no matter the extra degrees of freedom are considered or not. We use 26 discrete velocities to formulate the DBM
in Navier-Stokes level.

Besides recovering the hydrodynamic equations in the continuum limit, two
key points for a DBM are as below: (i) in terms of the
nonconserved moments, we can define two sets of measures for the deviation
of the system from its thermodynamic equilibrium state,
(ii) in the regimes
where the system deviates from its thermodynamic equilibrium, the DBM
may present more reasonable density, flow velocity and temperature than the
corresponding Navier-Stokes model only if the second order terms in Knudsen number in the Chapman-Enskog expansion are taken into account in the model construction. With the DBM we can study
\emph{simultaneously} both the hydrodynamic and thermodynamic behaviors.
Since the inverse of the transformation matrix $\mathbf{C}$ connecting the
discrete equilibrium distribution function $\mathbf{f}^{eq}$ and
corresponding moments $\mathbf{\hat{f}}^{eq}$ has been fixed, the extension
to multiple-relaxation-time DBM\cite{XuChen2014FoP} is
straightforward.
As for the DBM in spherical coordinates, if consider flow behaviors near the spherical center, the ``force term" (the term for geometric effects) should consider the higher order nonequilibrium effects. Specifically, $f = f^{eq}$ should be replaced by $f = f^{eq} + f^{(1)}$, even $f = f^{eq} + f^{(1)} + f^{(2)}$, etc. in the ``force term".
It should also be pointed out that, fixing the transformation matrix $%
\mathbf{C}$ and its inverse is only one of the possible schemes to get a
solution for the discrete equilibrium distribution function $\mathbf{f}^{eq}$%
. A second way to find a solution for the discrete equilibrium distribution
function $\mathbf{f}^{eq}$ is to follow the ideas used in Refs. \cite%
{Watari2004}. A difference is that the scheme
introduced in this work needs the minimum number of discrete velocities.

\section*{Acknowledgements}

The authors would like to thank Drs. Chuandong Lin, Yanbiao Gan and Feng Chen for helpful discussions. The work is supported by National Natural Science Foundation of China [under grant nos. 11475028,11772064 and U1530261] and Science Challenge Project (under Grant No. JCKY2016212A501 and TZ2016002).

\end{document}